\documentclass{bmvc2k}

\title{Rethinking Transfer Learning for Medical Image Classification}

\addauthor{Le Peng}{https://lepeng.org}{1}
\addauthor{Hengyue Liang}{https://hengyuel.github.io}{2}
\addauthor{Gaoxiang Luo}{https://gaoxiangluo.github.io}{1}
\addauthor{Taihui Li}{https://taihui.github.io}{1}
\addauthor{Ju Sun}{https://sunju.org}{1}

\addinstitution{
 Department of Computer Science and Engineering\\
 University of Minnesota\\
 Minneapolis, USA
}
\addinstitution{
  Department of Electrical and Computer Engineering\\
 University of Minnesota\\
 Minneapolis, USA
}

\runninghead{Peng ET AL.}{Rethinking Transfer Learning for MIC}

\usepackage{wrapfig}
\usepackage{booktabs}
\usepackage{multirow}
\usepackage{makecell}
\usepackage{graphicx}
\usepackage[export]{adjustbox}
\newcommand{\tabincell}[2]{\begin{tabular}{@{}#1@{}}#2\end{tabular}}
\usepackage{threeparttable}
\usepackage{rotating}
\usepackage{tablefootnote}
\usepackage{booktabs}
\usepackage{hyperref}

\usepackage{graphicx,amsfonts,amscd,amssymb,bm,url,color,latexsym}
\usepackage{physics}
\allowdisplaybreaks
\usepackage[capitalize,nameinlink]{cleveref}

\renewcommand{\mathbf}{\boldsymbol}

\newcommand{\mb}{\mathbf}
\newcommand{\mc}{\mathcal}

\newcommand{\bb}{\mathbb}

\newcommand{\set}[1]{\left\{ #1 \right\}}

\newcommand{\reals}{\bb R}

\usepackage{graphicx,amsfonts,amscd,amssymb,bm,url,color,latexsym}

\newcommand{\R}{\reals}

\newcommand{\paren}{\pqty}

\DeclareMathOperator{\st}{s.t.}

\newcommand{\T}{\intercal}

\newcommand{\expect}[1]{\bb E\left[ #1 \right]}

\usepackage{pifont}
\newcommand{\cmark}{\ding{51}}%
\newcommand{\xmark}{\ding{55}}%
\newcommand{\done}{\rlap{$\square$}{\raisebox{2pt}{\large\hspace{1pt}\cmark}}%
\hspace{-2.5pt}}
\newcommand{\tobedone}{\rlap{$\square$}{\large\hspace{1mm}\xmark}}

\usepackage{xcolor}
\usepackage{multirow}
\usepackage[colorlinks]{hyperref}
\usepackage{kantlipsum}

\begin{document}

\maketitle

\begin{abstract}
Transfer learning (TL) from pretrained deep models is a standard practice in modern medical image classification (MIC). However, what levels of features to be reused are problem-dependent, and uniformly finetuning all layers of pretrained models may be suboptimal. This insight has partly motivated the recent \emph{differential} TL strategies, such as TransFusion (TF) and layer-wise finetuning (LWFT), which treat the layers in the pretrained models differentially. In this paper, we add one more strategy into this family, called \emph{TruncatedTL}, which reuses and finetunes appropriate bottom layers and directly discards the remaining layers. This yields not only superior MIC performance but also compact models for efficient inference, compared to other differential TL methods.  Our code is available at: \url{https://github.com/sun-umn/TTL}.
\end{abstract}


\section{Introduction}\label{sec:intro}

Transfer learning (TL) is a common practice for medical image classification (MIC), especially when training data are limited. In typical TL pipelines for MIC, deep convolutional neural networks (DCNNs) pretrained on large-scale \emph{source tasks} (e.g., object recognition on ImageNet~\citep{5206848}) are finetuned as backbone models for \emph{target MIC tasks}; see, e.g., \cite{shin2016deep,van2016combining,rehman2020deep,huynh2016digital,antropova2017deep,ghafoorian2017transfer}, for examples of prior successes.

\begin{wrapfigure}{R}{0.4\linewidth}
    \centering
    \includegraphics[width=5.2cm]{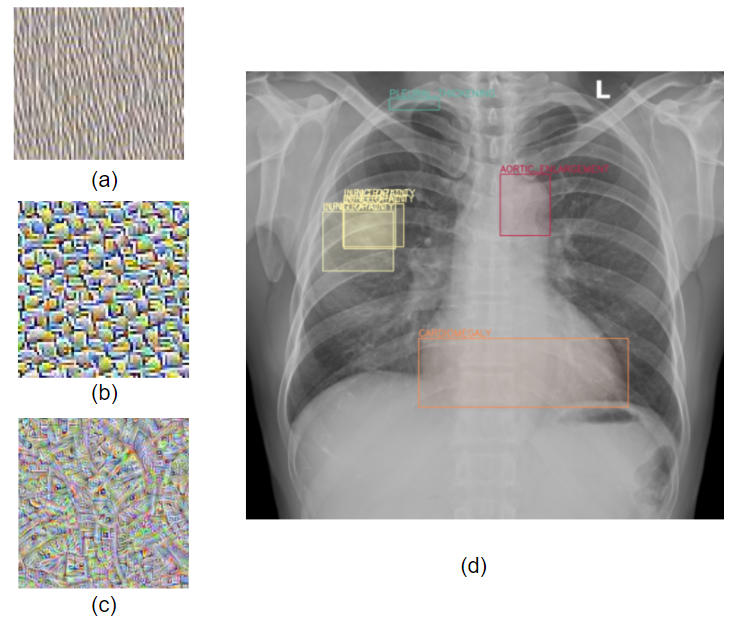}
    \caption{(left) The feature hierarchy learned by typical DCNNs, see {\color{red}Appendix E} for details; (right) Examples of diseases in a chest x-ray \cite{NguyenEtAl2021VinBigData}.}
    \label{fig:intro}
\end{wrapfigure}

The key to TL is feature reuse from the source to the target tasks, which leads to practical benefits such as fast convergence in training, and good test performance even if the target data are scarce~\citep{yosinski2014transferable}. Pretrained DCNNs extract increasingly more abstract visual features from bottom to top layers: from low-level corners and textures, to mid-level blobs and parts, and finally to high-level shapes and patterns~\citep{zeiler2014visualizing, yosinski2014transferable}. While shapes and patterns are crucial for recognizing and segmenting generic visual objects (see \cref{fig:intro} (left)), they are not necessarily the defining features for diseases: diseases can often take the form of abnormal textures and blobs, which correspond to low- to mid-level features (see \cref{fig:intro} (right)). \emph{So intuitively for MIC, we may only need to finetune a reasonable number of the bottom layers commensurate with the levels of features needed, and ignore the top layers.} However, standard TL practice for MIC retains all layers, and uses them as fixed feature extractors or finetunes them uniformly.

\citet{tajbakhsh2016convolutional,NEURIPS2019_eb1e7832} depart from the uniform TL approach and propose TL methods that treat top and bottom layers differently. Prioritizing high-level features and the classifier, \citet{tajbakhsh2016convolutional} proposes \emph{layer-wise finetuning} (LWFT) that finetunes an appropriate number of top layers and freezes the remaining bottom layers. In comparison, to improve training speed while preserving performance, \citet{NEURIPS2019_eb1e7832} proposes \emph{TransFusion} (TF) that finetunes bottom layers but retrains a coarsened version of top layers from scratch. 

Neither of the above \emph{differential} TL strategies clearly address the conundrum of why top layers are needed when only the features in bottom layers are to be reused. To bridge the gap, in this paper, we propose a novel, perhaps radical, TL strategy: remove top layers after an appropriate cutoff point, and finetune the truncated model left, dubbed \emph{TruncatedTL} (TTL)---this is entirely consistent with our intuition about the feature hierarchy. Our main contributions include: 1) \textbf{confirming the deficiency of full TL}. By experimenting with full and differential TL strategies---including our TTL---on three MIC tasks and three popularly used DCNN models for MIC, we find that full TL (FTL) is almost always suboptimal in terms of classification performance, confirming the observation in \citet{tajbakhsh2016convolutional}. 2) \textbf{proposing TruncatedTL (TTL) that leads to effective and compact models}. Our TTL outperforms other differential TL methods, while the resulting models are always smaller, sometimes substantially so. This leads to reduced computation and fast prediction during inference, and can be particularly valuable when dealing with 3D medical data such as CT and MRI images. 3) \textbf{quantifying feature transferability in TL for MIC}. We use singular vector canonical correlation analysis (SVCCA)~\citep{raghu2017svcca} to analyze feature transferability and confirm the importance of low- and mid-level features for MIC. The quantitative analysis also provides insights on how to choose high-quality truncation points to further optimize the model performance and efficiency of TTL.

\section{Related Work}\label{sec:realted_works}

\textbf{Deep TL} TL by reusing and finetuning visual features learned in DCNNs entered computer vision (CV) once DCNNs became the cornerstone of state-of-the-art (SOTA) object recognition models back to 2012. For example, \citet{donahue2014decaf, sharif2014cnn} propose using part of or full pretrained DCNNs as feature extractors for generic visual recognition. \citet{yosinski2014transferable} studies the hierarchy of DCNN-based visual features, characterizes their transferability, and proposes finetuning pretrained features to boost performance on target tasks. Moreover, \citet{erhan2009visualizing,zeiler2014visualizing} propose techniques to visualize visual features and their hierarchy. This popular line of TL techniques is among the broad family of TL methods for knowledge transfer from source to target tasks based on deep networks~\citep{tan2018survey} and other learning models~\citep{9134370}, and is the focus of this paper. 

\begin{wrapfigure}{R}{0.4\linewidth}
    \centering
    \vspace{-0.3cm}
    \includegraphics[width=5cm]{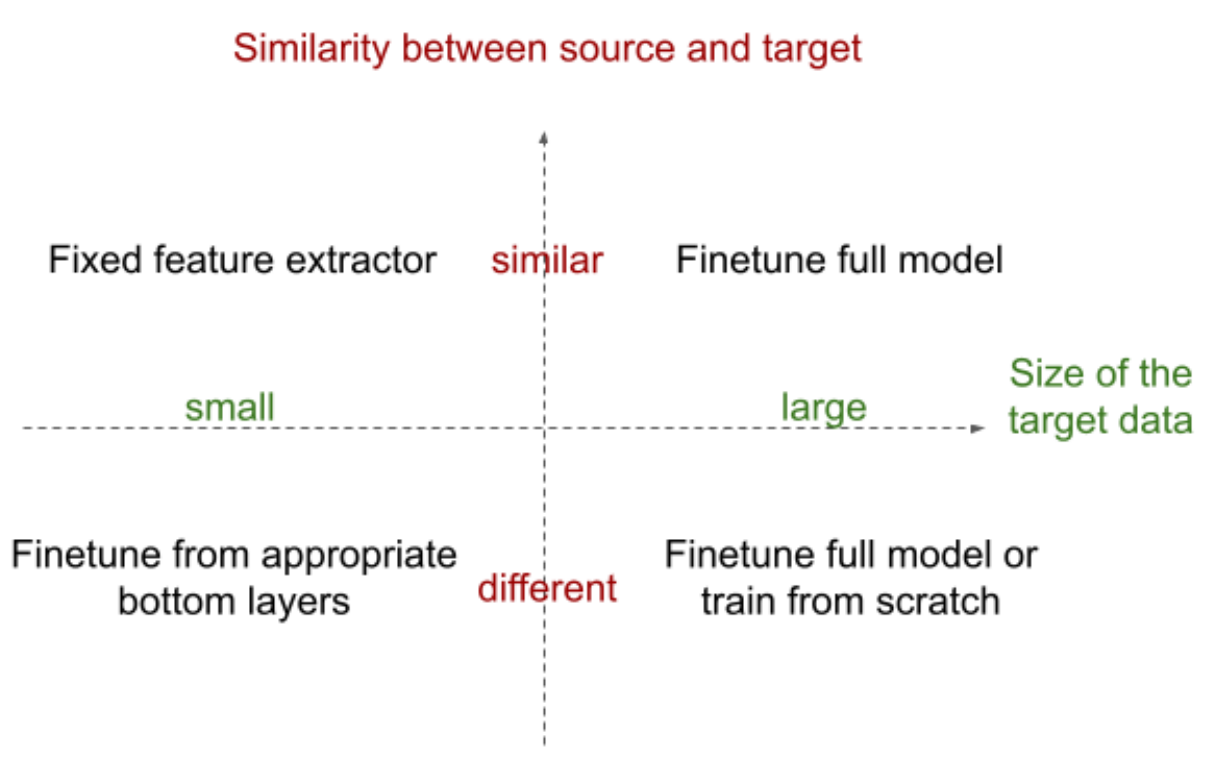}
    \vspace{-0.3cm}
    \caption{TL strategies and their usage scenarios}
    \vspace{-0.33cm}
    \label{fig:rela}
\end{wrapfigure}

\noindent\textbf{General TL stategies} While pretrained DCNNs can be either used as fixed feature extractors, or partially or fully finetuned on the target data, \citet{yosinski2014transferable} argues that bottom layers learn generic features while top layers learn task-specific features, leading to folklore guidelines on how to choose appropriate TL strategies in various scenarios, as summarized in \cref{fig:rela}. 

Our intuition about the feature hierarchy is slightly different: bottom layers learn low-level features that are spatially localized, and top layers learn high-level features that are spatially extensive (see \cref{fig:intro}). Although the two hierarchies may be aligned for most cases, they are distinguished by whether spatial scales are considered: task-specific features may be spatially localized, e.g., texture features to classify skin lesions~\citep{tschandl2018ham10000}, and general features may be spatially extensive, e.g., generic brain silhouettes in brain MRI images. Moreover, the spatial-scale hierarchy is built into DCNNs by design~\citep{bronstein2021geometric}. So we argue that the intuition about the low-high spatial feature hierarchy is more pertinent. We note that none of the popular strategies as summarized in \cref{fig:rela} modify the pretrained DCNN models (except for the final multi-layer perceptron, MLP, classifiers)---in contrast to TF and our TTL.

\noindent\textbf{Differential TL} Early work on TL for MIC~\citep{shin2016deep,van2016combining,rehman2020deep,huynh2016digital,antropova2017deep,hosny2018skin,sun2021prospective} and segmentation~\citep{van2014transfer,ronneberger2015u,ghafoorian2017transfer,jiang2018retinal, karimi2021transfer} parallels the relevant developments in CV, and mostly uses DCNNs pretrained on CV tasks as feature extractors or initializers (i.e., for finetuning). In fact, these two strategies remain dominant according to the very recent survey~\citep{Kim2022} on TL for MIC, which reviews around $120$ relevant papers. But the former seems inappropriate as medical data are disparate from natural images dealt with in CV. From the bottom panel of \cref{fig:rela}, finetuning at least part of the DCNNs is probably more competitive even if the target data are limited. In this line, LWFT~\citep{tajbakhsh2016convolutional} finetunes top layers and freeze bottom layers, and incrementally allows finetuning more layers during model selection---this is  recommended in \citet{Kim2022} as a practical TL strategy for MIC that strikes a balance between training efficiency and performance. Similarly, TF \citep{NEURIPS2019_eb1e7832} coarsens top layers which are then trained from scratch, and finetunes bottom layers from pretrained weights. Both LWFT and TF take inspiration from the general-specific feature hierarchy. In contrast, motivated by the low-high spatial feature hierarchy, our novel TTL method removes the top layers entirely and directly finetunes the truncated models. Our experiments in \cref{sec:application} confirm that TTL surpasses LWFT and TF with improved performance and reduced inference cost. 

\noindent\textbf{Compact models for MIC} Both TF and our TTL lead to reduced models that can boost the inference efficiency, the first of its kind in TL for MIC, although the TF paper~\citep{NEURIPS2019_eb1e7832} does not stress this point. Compact models have been designed for specific MIC tasks, e.g., \citet{NEURIPS2019_eb1e7832,huang2020penet}, but our evaluation on the task of~\citet{huang2020penet} in \cref{subsec:PE} suggests that the differential TL strategies based on generic pretrained models, particularly our TTL, can outperform TL based on handcrafted models. Moreover, the growing set of methods for model quantization and compression~\citep{wang2020towards, krishnamoorthi2018quantizing,polino2018model, han2015deep} are equally applicable to both the original models and the reduced models.

\section{Efficient transfer learning for MIC}\label{sec:method}
Let $\mc X \times \mc Y$ denote any input-output (or feature-label) product space, and $\mc D_{\mc X \times \mc Y}$ a distribution on $\mc X \times \mc Y$. TL considers a source task $\mc D_{\mc X_s \times \mc Y_s} \mapsto p_s$, where $p_s$ is a desired predictor, and a target task $\mc D_{\mc X_t \times \mc Y_t} \mapsto p_t$. In typical TL, $\mc X_t \times \mc Y_t$ may be different from $\mc X_s \times \mc Y_s$, or at least $\mc D_{\mc X_t \times \mc Y_t} \ne \mc D_{\mc X_s \times \mc Y_s}$ even if $\mc X_t \times \mc Y_t = \mc X_s \times \mc Y_s$. The goal of TL is to transfer the knowledge from the source task $\mc D_{\mc X_s \times \mc Y_s} \mapsto p_s$ that is solved beforehand to the target task $\mc D_{\mc X_t \times \mc Y_t} \mapsto p_t$~\citep{pan2009survey,tan2018survey}. 

In this paper, we restrict TL to reusing and finetuning pretrained DCNNs for MIC. In this context, the source predictor $p_s = h_s \circ f_L \circ \dots \circ f_1$ is pretrained on a large-scale source dataset $\set{(x_i, y_i)} \sim_{iid} \mc D_{\mc X_s \times \mc Y_s}$. Here, the $f_i$'s are $L$ convolulational layers, and $h_s$ is the final MLP classifier. To perform TL, $h_s$ is replaced by a new MLP predictor $h_t$ with prediction heads matching the target task (i.e., with the desired number of outputs) to form the new model $p_t = h_t \circ f_L \circ \dots \circ f_1$. Two dominant approaches of TL for MIC are: 1) \emph{fixed feature extraction}: freeze the pretrained weights of $f_L \circ \dots \circ f_1$, and optimize $h_t$ from random initialization so that $p_t$ fits the target data; 2) \emph{full transfer learning (FTL)}:  optimize all of $h_t \circ f_L \circ \dots \circ f_1$, with $h_t$ from random initialization whereas $f_L \circ \dots \circ f_1$ from their pretrained weights so that $p_t$ fits the target data.

\begin{figure}
\begin{tabular}{cc}
\bmvaHangBox{\hspace{-2.24mm}\fbox{\hspace{-2.24mm}\includegraphics[width=6.9cm]{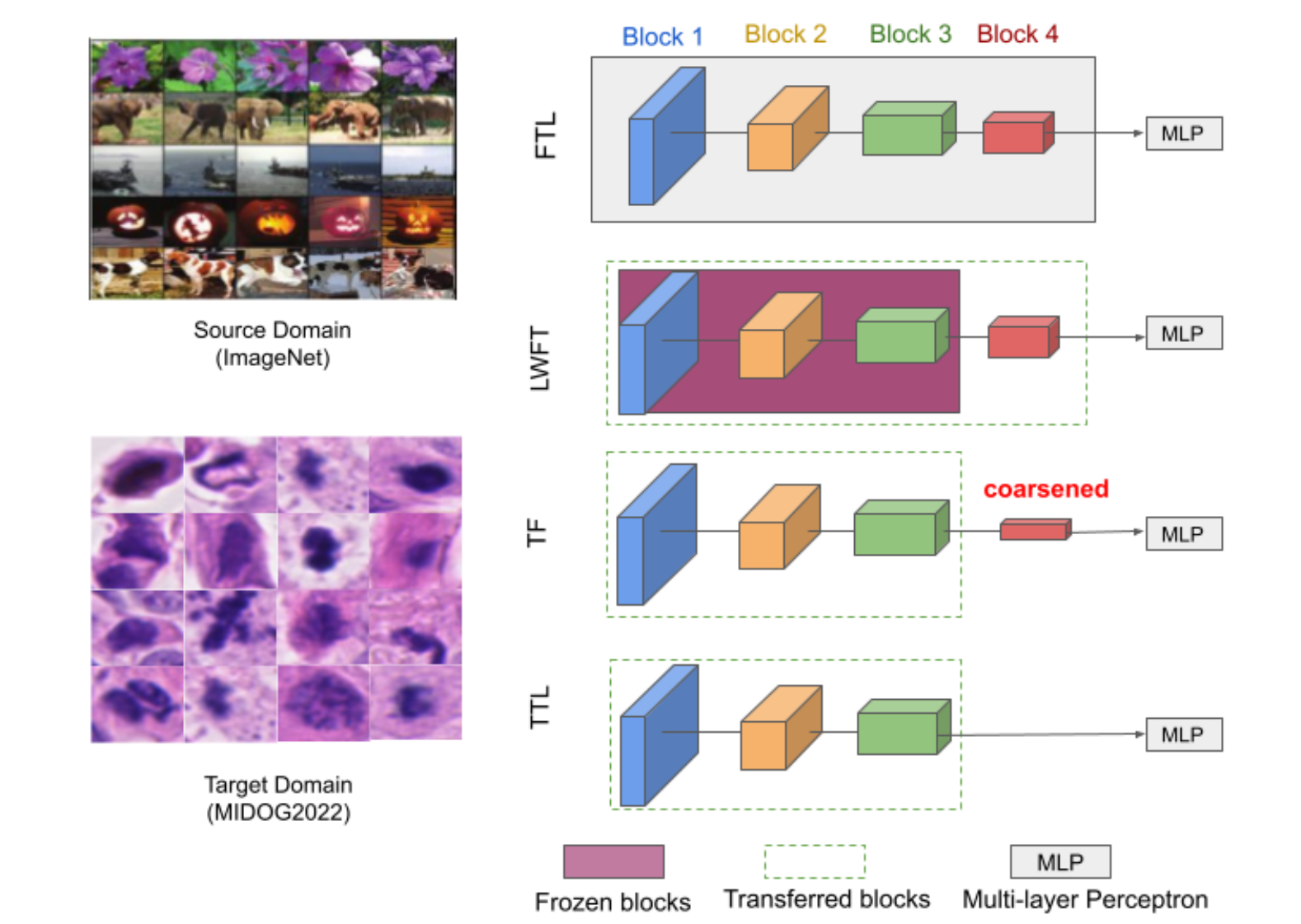}}}&
\bmvaHangBox{\hspace{-3.24mm}\fbox{\parbox{5.6cm}{\includegraphics[width=5.3cm]{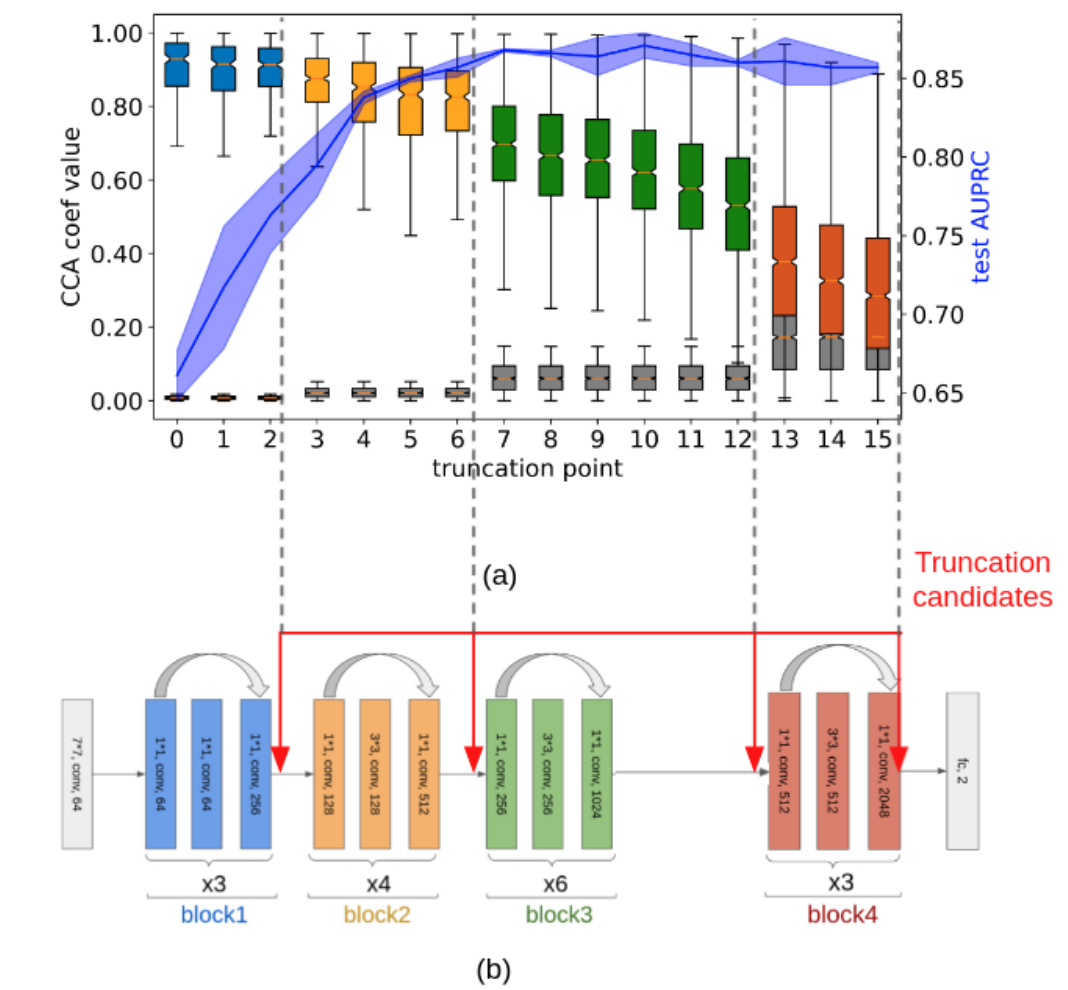}}}}\\
(i)&(ii)
\end{tabular}
\caption{\textbf{(i)} Overview of typical TL setup, and the four TL methods that we focus on in this paper. \textbf{(ii)} Illustration of feature transferability and the performance of different levels of features on BIMCV. We take the ResNet50 model pretrained on ImageNet, and perform a full TL on BIMCV. We consider $17$ natural truncation/cutoff points that do not cut through the skip connections.}
\label{fig:overview}
\vspace{-0.5cm}
\end{figure}

\subsection{Prior differential TL approaches}
The two differential TL methods for MIC, i.e., LWFT~\citep{tajbakhsh2016convolutional} and TF~\citep{raghu2019direct}, differ from the dominant TL approaches in that they treat the top and bottom layers differently, as illustrated in \cref{fig:overview} (i). 

\noindent\textbf{Layer-wise finetuning}
LWFT does not distinguish MLP layers and convolutional layers. So slightly abusing our notation, assume that the pretrained DCNN is $f_N \circ f_{N-1} \circ \dots \circ f_1$, where $N$ is the total number of layers including both the MLP and convolutional layers. LWFT finetunes the top $k$ layers $f_N \circ f_{N-1} \circ \dots \circ f_{N-k+1}$, and freezes the bottom $N-k$ layers $f_{N-k} \circ \dots \circ f_{1}$. The top layer $f_N$ is finetuned with a base learning rate (LR) $\eta$, and the other $k-1$ layers with a LR $\eta/10$. To find an appropriate $k$, \citet{tajbakhsh2016convolutional} proposes an incremental model selection procedure: start with $k=1$ layer, and include one more layer into finetuning if the previous set of layers does not achieve the desired level of performance\footnote{Finetuning starts with the original pretrained weights each time. }. Although LWFT was originally proposed for AlexNet~\citep{krizhevsky2012imagenet}, it can be generalized to work with advanced DCNN models such as ResNets and DenseNets that have block structures.

\noindent\textbf{TransFusion}
TF reuses bottom layers while slimming down top layers. Formally, for a cutoff index $k$, the pretrained model $p_s \circ f_L \circ \dots \circ f_{k+1} \circ f_k \dots \circ f_1$ is replaced by $p_t \circ f_L^{HV} \circ \dots \circ f_{k+1}^{HV} \circ f_k \dots \circ f_1$, where HV means halving the number of channels in the designated layer. TF then trains the coarsened model on the target data with the first half $f_k \dots \circ f_1$ initialized by the pretrained weights (i.e., finetuning) and $p_t \circ f_L^{HV} \circ \dots \circ f_{k+1}^{HV}$ initialized by random weights (i.e., training from scratch). The cutoff point $k$ is the key hyperparameter in TF. TF was originally proposed to boost the finetuning speed, but we find that it often also boosts the classification performance compared to FTL; see \cref{sec:application}. 

\subsection{Our truncated TL approach}
\label{sec:ttl_method}
Our method is radically simple: for an appropriate cutoff index $k$, we take the $k$ bottom layers $f_{k} \circ \dots \circ f_1$ from the pretrained DCNN model and then form and finetune the new predictor $h_t \circ f_{k} \circ \dots \circ f_1$. Our TruncatedTL (TTL) method is illustrated in \cref{fig:overview}(i). 

The only crucial hyperparameter for TTL is the cutoff index $k$, which depends on the DCNN model and problem under consideration. For SOTA ResNet and DenseNet models that are popularly used in TL for MIC, there are always $4$ convolutional blocks each consisting of repeated basic convolutional structures; see, e.g., the bottom of \cref{fig:overview} (ii) for an illustration of ResNet50. So we propose a hierarchical search strategy: \textbf{(1) Stage 1: coarse block search}. Take the block cutoffs as candidate cutoff points, and report the best-performing one; \textbf{(2) Stage 2: fine-grained layer search}. Search over the neighboring layers of the cutoff from \textbf{Stage 1} to optimize the performance. Since $k$ is also a crucial algorithm hyperparameter for TF and LWFT, we also adopt the same hierarchical search strategy for them when comparing the performance.

To quickly confirm the efficiency of TTL, we apply TTL and competing TL methods on an X-ray-based COVID-19 classification task on the BIMCV dataset~\citep{vaya2020bimcv}; more details about the setup can be found in \cref{subsec:bimcv}. We pick the COVID example, as the salient radiological patterns in COVID X-rays such as multifocal and bilateral ground glass opacities and consolidations are low- to mid-level visual features~\citep{shi2020radiological} and hence we can easily see the benefit of differential TL methods including our TTL. We measure the classification performance by both AUROC (area under the receiver-operating-characteristic curve) and AUPRC (area under the precision-recall curve), and measure the inference complexity by P (number of parameters in the model, M---millions), MACs (multiply-add operation counts~\citep{ptflops}\footnote{\url{https://github.com/sovrasov/flops-counter.pytorch}}, G---billion), T (wallclock run time in milliseconds; details of our computing environment can be found supplementary material). 

\cref{tab:bimcv} summarizes the results, and we observe that: (1) differential TL methods (TF, LWFT, and TTL) perform better or at least on par with FTL, and the layer-wise search in the second stage further boosts their performance. Also, TTL is the best-performing TL method with the two-stage hierarchical search; (2) TF and TTL that slim down the model lead to reduced model complexity and hence considerably less run time. TTL is a clear winner in terms of both performance and inference complexity. (3) TTL tends to focus on the foreground lesion area instead of the background told by the Grad-CAM visualization.

\begin{table}

\resizebox{0.5\columnwidth}{!}{%
    \def\arraystretch{1.3}
    \addtolength{\tabcolsep}{0pt}
  \begin{tabular}{c |c c c c c}
  \Xhline{3\arrayrulewidth}
  Method&
  AUROC$\uparrow$&
  AUPRC$\uparrow$ &
  P(M)$\downarrow$ &
  MACs(G)$\downarrow$ &
  T(ms)$\downarrow$ \\ 
  \toprule
  \midrule
  {FTL} &\tabincell{c}{{$0.849 \pm 0.001$}} & \tabincell{c}{{$0.857 \pm 0.003$}} & \tabincell{c}{{23.5}} & \tabincell{c}{{4.12}} & \tabincell{c}{{3.59}} \\
  \hline
  {TF-1} & \tabincell{c}{{$0.856 \pm 0.011$}} & \tabincell{c}{{$0.863 \pm 0.012$}} & \tabincell{c}{12.9} & \tabincell{c}{3.56} & \tabincell{c}{{3.55}} \\
  {LWFT-1} & \tabincell{c}{{$0.848 \pm 0.002$}} & \tabincell{c}{{{$0.861 \pm 0.004$}}} & \tabincell{c}{{23.5}} & \tabincell{c}{4.12} & \tabincell{c}{3.59} \\
  {\textbf{TTL-1}} & \tabincell{c}{$0.851 \pm 0.002$} & \tabincell{c}{$0.860 \pm 0.002$} & \tabincell{c}{8.55} & \tabincell{c}{3.31} & \tabincell{c}{3.19} \\
  \midrule
  {TF-2} & \tabincell{c}{$0.856 \pm 0.011$} & \tabincell{c}{$0.863 \pm 0.012$} & \tabincell{c}{12.9} & \tabincell{c}{3.56} & \tabincell{c}{{3.56}} \\
  {LWFT-2} & \tabincell{c}{{{$0.853 \pm 0.005$}}} & \tabincell{c}{{$0.861 \pm 0.001$}} & \tabincell{c}{{23.5}} & \tabincell{c}{{4.12}} & \tabincell{c}{3.56} \\
  {\textbf{TTL-2}} & \tabincell{c}{\textcolor{red}{{$\mathbf{0.861 \pm 0.013}$}}} & \tabincell{c}{\textcolor{red}{$\mathbf{0.871 \pm 0.008}$}} & \tabincell{c}{\textcolor{red}{\textbf{6.31}}} & \tabincell{c}{\textcolor{red}{\textbf{2.87}}} & \tabincell{c}{\textcolor{red}{\textbf{2.97}}} \\
  
  \Xhline{3\arrayrulewidth}
  \end{tabular}
}
\begin{tabular}{c}
     \includegraphics[width=6.2cm]{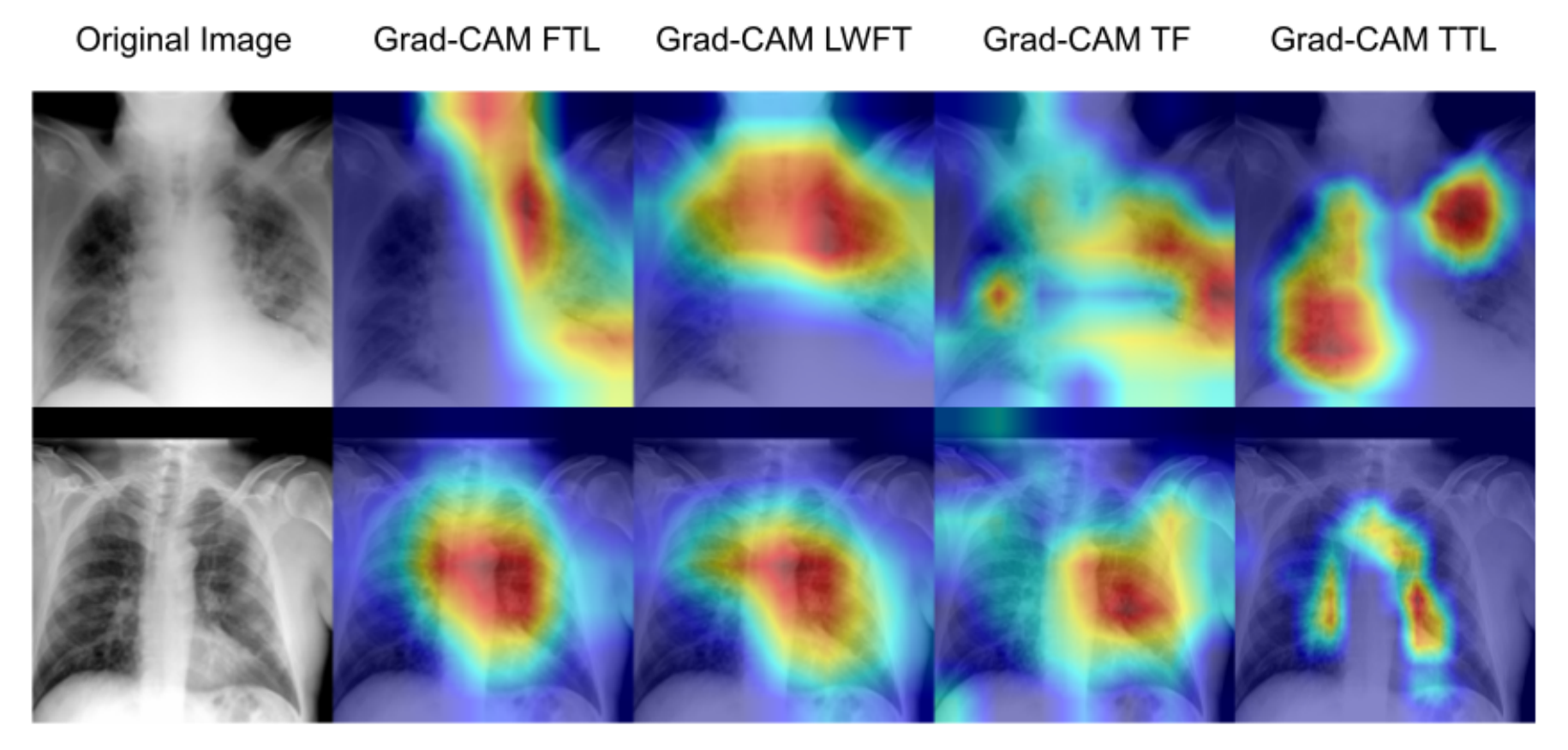}
\end{tabular}
\caption{A comparison of different TL strategies on BIMCV COVID-19 diagnosis task. (left) a summary of classification performance, (right) visualizations using Grad-CAM}
\label{tab:bimcv}
\vspace{-0.5cm}
\end{table}

\subsection{Transferablity analysis}
Besides the positive confirmation above, in this section, we provide quantitative corroboration for our claim that top layers might not be needed in TL for MIC. To this end, we need a variant of the classical canonical correlation analysis (CCA), singular-vector CCA (SVCCA)~\citep{raghu2017svcca}. 

\noindent\textbf{SVCCA for quantifying feature correlations}
CCA is a classical statistical tool for measuring the linear correlation between random vectors. Suppose that $\mb x \in \R^p$ and $\mb y \in \R^q$ are two random vectors containing $p$ and $q$ features, respectively. CCA seeks the linear combinations $\mb u^\T \mb x$ and $\mb v^\T \mb y$ of the two sets of features with the largest covariance $\mathrm{cov}(\mb u^\T \mb x, \mb v^\T \mb y)$. Assume $\expect{\mb x} = \mb 0$ and $\expect{\mb y} = \mb 0$. The problem can be formulated as 
\begin{align*} \label{eq:cca_1st_stat}
  \max_{\mb u, \mb v} \; \mb u^\T \mb \Sigma_{\mb x\mb y} \mb v  \quad \st\; \mb u^\T \mb \Sigma_{\mb x\mb x} \mb u = 1, \; \mb v^\T \mb \Sigma_{\mb y\mb y} \mb v = 1, 
\end{align*}
where $\mb \Sigma_{\mb x\mb y} \doteq \expect{\mb x \mb y^\T}$, $\mb \Sigma_{\mb x\mb x} \doteq \expect{\mb x \mb x^\T}$, and $\mb \Sigma_{\mb y\mb y} \doteq \expect{\mb y \mb y^\T}$,\footnote{$\doteq$ means ``defined as".} and the constraints fix the scales of $\mb u$ and $\mb v$ so that the objective does not blow up. Once the $(\mb u, \mb v)$ pair with the largest covariance is computed, subsequent pairs are computed iteratively in a similar fashion with the additional constraint that new linear combinations are statistically decorrelated with the ones already computed. The overall iterative process can be written compactly as 
\begin{equation*}
  \begin{aligned}
    \max_{\mb U \in \R^{p \times k}, \mb V \in \R^{q \times k}} \; & \trace\paren{\mb U^\T \mb \Sigma_{\mb x \mb y}\mb V} \quad
    \st\; & \mb U^\T \mb \Sigma_{\mb x \mb x}\mb U = \mb I_{k \times k}, \; \mb V^\T \mb \Sigma_{\mb y \mb y}\mb V = \mb I_{k \times k}, 
  \end{aligned}
\end{equation*}
which computes the first $k$ pairs of most correlated linear combinations. In practice, all the covariance matrices $\mb \Sigma_{\mb x \mb y}$, $\mb \Sigma_{\mb x \mb x}$, and $\mb \Sigma_{\mb y \mb y}$ are replaced by their finite-sample approximations, and the covariances of the top $k$ most correlated pairs are the top $k$ singular values of $\mb \Sigma^{-1/2}_{\mb x \mb x} \mb \Sigma_{\mb x \mb y} \mb \Sigma^{-1/2}_{\mb y \mb y}$~\citep{UurtioEtAl2018Tutorial}, all of which lie in $[0, 1]$. We call these singular values the \emph{CCA coefficients}. Obviously, high values in these coefficients indicate high levels of correlation.

For our subsequent analyses, we typically need to find the correlation between the features of two data matrices $\mb X \in \R^{n \times p}$ and $\mb Y \in \R^{n \times q}$, where $n$ is the number of data points. SVCCA performs principal component analysis (PCA) separately on $\mb X$ and $\mb Y$ first, so that potential noise in the data is suppressed and the ensuing CCA analysis becomes more robust. We typically plot the CCA coefficients in descending order when analyzing two group of features.

\begin{figure}
\begin{tabular}{cc}
\bmvaHangBox{\hspace{-1.24mm}\fbox{\hspace{-1.24mm}\includegraphics[width=8.0cm]{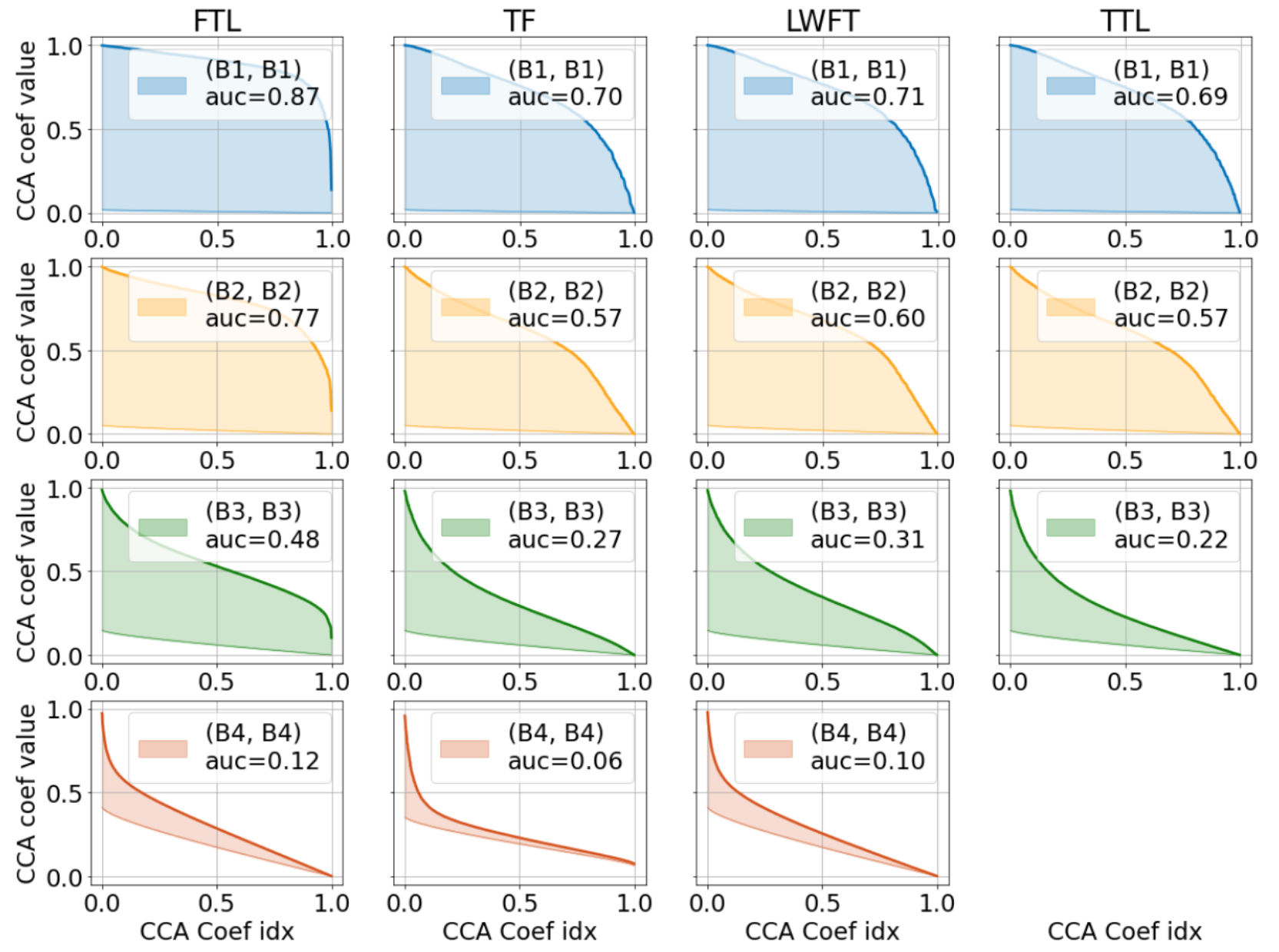}}}&
\bmvaHangBox{\hspace{-3.24mm}\fbox{\parbox{4.2cm}{\includegraphics[width=3.9cm]{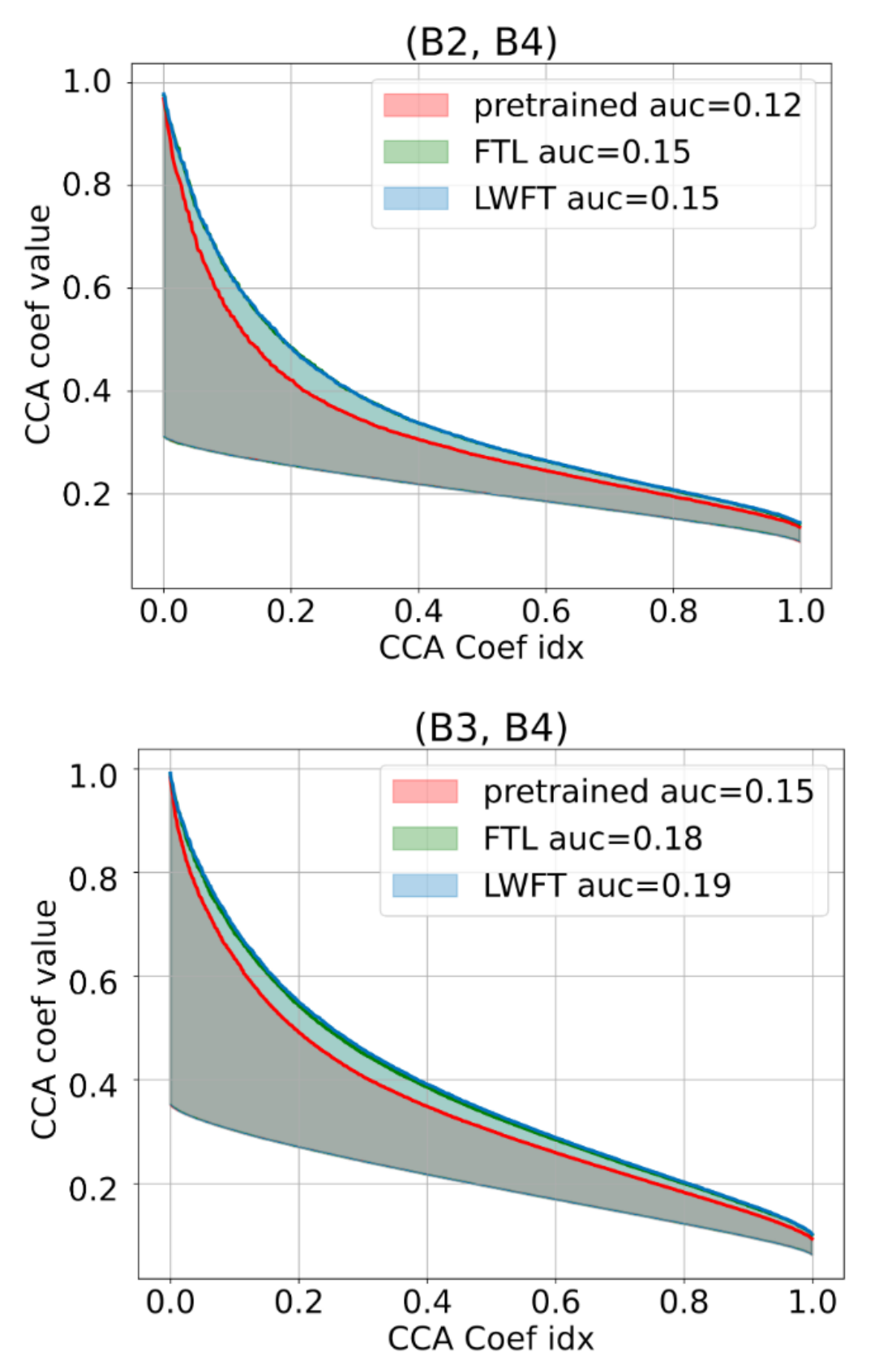}}\hspace{-2.24mm}}}\\
(i)&(ii)
\end{tabular}
\caption{SVCCA on COVID-19 diagnosis task. Bold curve indicates the CCA coefficients for learned features, while light curve indicates the correlation for two uncorrelated random features. So the area between the two curves is a quantitative measure of the correlation between the said blocks. We normalize all the indices of the CCA coefficients to be $[0, 1]$. We compare (i) features learned from trained and finetuned model at the same layer, and (ii) features learned in the finetuned model but different blocks
    }
\label{fig:svcca}
\vspace{-0.5cm}
\end{figure}

\noindent\textbf{Layer transferability analysis}
With SVCCA, we are now ready to present quantitative results to show that reusing and finetuning top layers may be unnecessary. We again take the BIMCV dataset for illustration. 

\textit{Features in top layers change substantially, but the changes do not help improve the performance} In \cref{fig:svcca} (i) we present the per-layer correlation of the learnt features before and after FTL. First, the correlation monotonically decreases from bottom to top layers, suggesting increasingly dramatic feature finetuning/learning. For features residing in block1 through block3, the correlation levels are substantially higher than that of random features, suggesting considerable feature reuse together with the finetuning. However, for features in block4 which contain the top layers, the correlation level approaches that of random features. So these high-level features are drastically changed during FTL and there is little reuse. The changes do not help and in fact hurt the performance: when we take intermediate features after the FTL and train a classifier based on each level of them, we find that the performance peaks at block3, and starts to degrade afterward. From \cref{fig:svcca} (i) (only per-block feature correlations are computed to save space), we find similar patterns in TF, LWFT, and our TTL also: features in bottom blocks are substantially more correlated than those in top blocks, and features in block4---which we remove in our TTL---are almost re-learned as their correlation with the original features come close to that between random features. 

\textit{Features in top layers become more correlated with bottom layers after TL} The ``horizontal'' analysis above says the features of the top layers are almost re-learned in FTL, LWFT, and TF, but it remains unclear what features are learned there. If we believe that high-level features are probably not useful for COVID classification~\citep{shi2020radiological}, a reasonable hypothesis is that these top layers actually learn features that are more correlated with those of lower layers after TL. This seems indeed the case, as shown in \cref{fig:svcca} (ii): the correlation level between the block2 and block 4 features, as well as between block3 and block4 features, increases both visibly and quantitatively. 

Given the above two sets of findings, our idea in TTL to remove the redundant top layers and keep the essential bottom layers is reasonable toward effective and compact models.

\section{Experiments} \label{sec:application}

\noindent\textbf{Experiment setup}  \label{sec:exp_setup}
We systematically compare FTL, TF, LWFT, and our TTL on $3$ MIC tasks covering both 2D and 3D image modalities, and also explore a 2D lung segmentation task.  For the 3D MIC task, we choose ResNeXt3D-101 as the default model and compare it with PENet which is a handcrafted model in~\citet{Huang2020}. Both models are first pretrained on the kinetics-600 dataset~\citep{carreira2018short}, and then finetuned with the same setting as in~\citet{Huang2020}: $0.01$ initial LR for randomly initialized weights and $0.1$ for pretrained weights, SGD (momentum $=0.9$) optimizer, cosine annealing LR scheduler, $100$ epochs of training, and best model selected based on the validation AUROC. For experiments involving randomness, we repeat them $3$ times and report the mean and standard deviation. More detailed experiment setup, ablation studies, and other explorations can be found in supplementary materials.

\begin{table}
\def\arraystretch{1.3}
\addtolength{\tabcolsep}{0pt}
\centering
\hspace{-0.45cm}
\quad
\resizebox{0.5\linewidth}{!}{%
\begin{tabular}{c | c c c c c}
\Xhline{3\arrayrulewidth}
Method&
AUROC$\uparrow$ &
AUPRC$\uparrow$ &
P(M)$\downarrow$ &
MACs(G)$\downarrow$ &
T(ms)$\downarrow$ \\
\midrule
\midrule

{FTL}
&\tabincell{c}{$0.925 \pm 0.003$}
&\tabincell{c}{$0.917 \pm 0.003$}
&\tabincell{c}{23.5}
&\tabincell{c}{4.12}
&\tabincell{c}{3.59}
\\
{TF-1}
&\tabincell{c}{$0.925 \pm 0.002$}
&\tabincell{c}{$0.918 \pm 0.002$}
&\tabincell{c}{12.9}
&\tabincell{c}{3.56}
&\tabincell{c}{3.50}
\\
{LWFT-1}
&\tabincell{c}{$0.920 \pm 0.003$}
&\tabincell{c}{$0.913 \pm 0.005$}
&\tabincell{c}{23.5}
&\tabincell{c}{4.12}
&\tabincell{c}{3.55}
\\
{\textbf{TTL-1}}
&\tabincell{c}{\textcolor{red}{{$\mathbf{0.928 \pm 0.004}$}}}
&\tabincell{c}{\textcolor{red}{{$\mathbf{0.921 \pm 0.003}$}}}
&\tabincell{c}{{\textcolor{red}{\textbf{8.55}}}}
&\tabincell{c}{{\textcolor{red}{\textbf{3.31}}}}
&\tabincell{c}{{\textcolor{red}{\textbf{2.99}}}}
\\

\midrule

{TF-2}
&\tabincell{c}{$0.925 \pm 0.002$}
&\tabincell{c}{$0.918 \pm 0.002$}
&\tabincell{c}{12.9}
&\tabincell{c}{3.56}
&\tabincell{c}{3.50}
\\
{LWFT-2}
&\tabincell{c}{$0.925 \pm 0.001$}
&\tabincell{c}{$0.919 \pm 0.001$}
&\tabincell{c}{23.5}
&\tabincell{c}{4.12}
&\tabincell{c}{3.56}
\\
{\textbf{TTL-2}}
&\tabincell{c}{\textcolor{red}{{$\mathbf{0.928 \pm 0.004}$}}}
&\tabincell{c}{\textcolor{red}{{$\mathbf{0.921 \pm 0.003}$}}}
&\tabincell{c}{{\textcolor{red}{\textbf{8.55}}}}
&\tabincell{c}{{\textcolor{red}{\textbf{3.31}}}}
&\tabincell{c}{{\textcolor{red}{\textbf{2.99}}}}

\\

\Xhline{3\arrayrulewidth}
\end{tabular}
}
\begin{tabular}{c}
     \includegraphics[width=5.8cm]{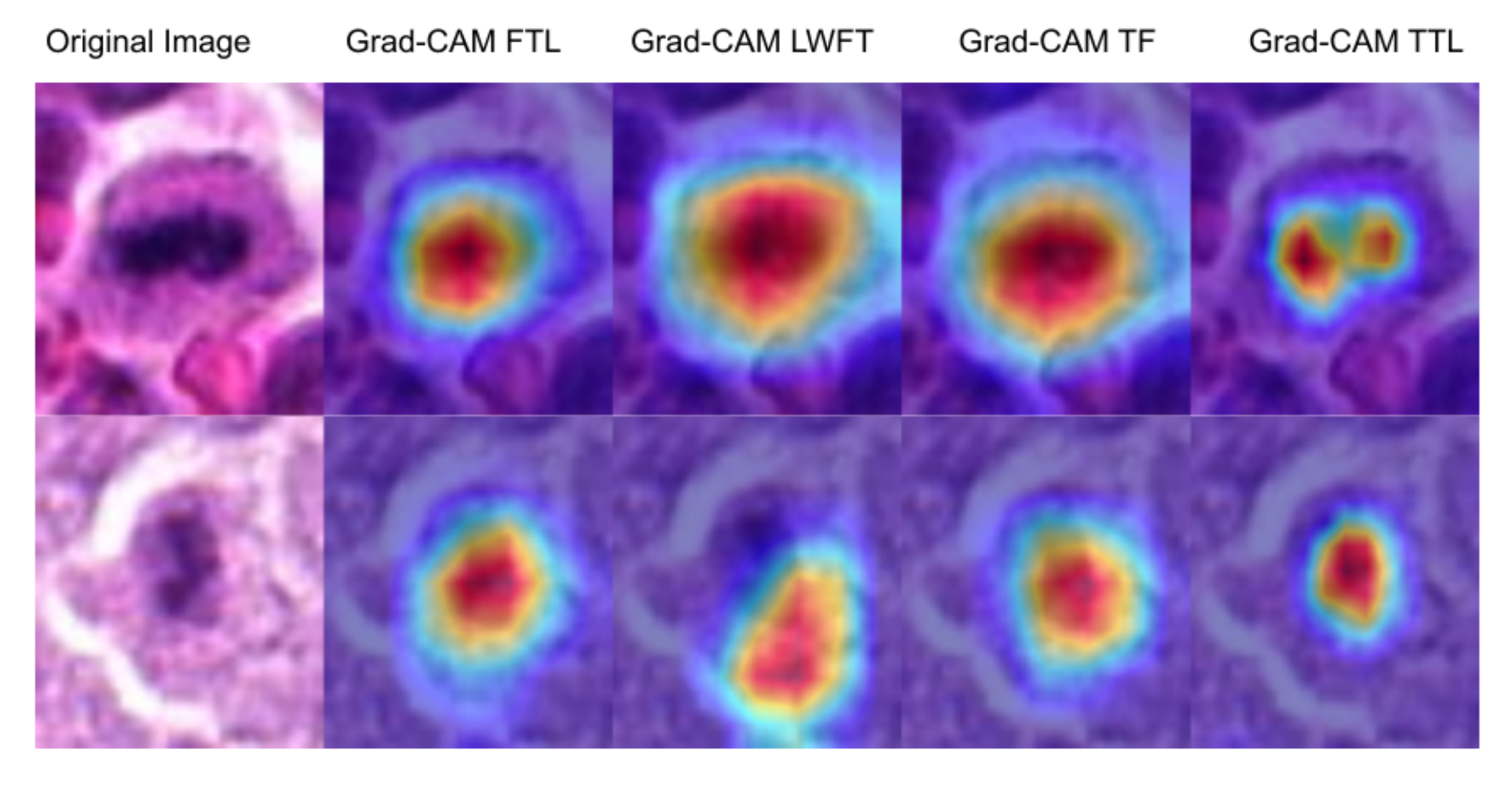}
\end{tabular}

\vspace{0.2cm}
\hspace{-0.45cm}
\resizebox{0.48\linewidth}{!}{%
\begin{tabular}{c | c c c c c}
\Xhline{3\arrayrulewidth}
Method&
AUROC$\uparrow$&
AUPRC$\uparrow$ &
P(M)$\downarrow$ &
MACs(G)$\downarrow$ &
T(\textcolor{blue}{s})$\downarrow$ \\
\midrule
\midrule

{PENet}
&\tabincell{c}{{$0.822 \pm 0.010$}}
&\tabincell{c}{{$0.855 \pm 0.007$}}
&\tabincell{c}{{28.4}}
&\tabincell{c}{\textcolor{red}{$\mb{51.7}$}}
&\tabincell{c}{\textcolor{red}{$\mb{1.59}$e-$\mb 2$}}
\\
\hline
{FTL}
    &\tabincell{c}{{$0.821 \pm 0.010$}}
    &\tabincell{c}{{$0.867 \pm 0.006$}}
    &\tabincell{c}{{47.5}}
    &\tabincell{c}{{66.3}}
    &\tabincell{c}{{1.96e-2}}
\\
    {TF-1}
    &\tabincell{c}{{{$0.849 \pm 0.020$}}}
    &\tabincell{c}{{$0.886 \pm 0.017$}}
    &\tabincell{c}{{36.1}}
    &\tabincell{c}{{64.9}}
    &\tabincell{c}{1.93e-2}
\\
{LWFT-1}
    &\tabincell{c}{$0.817 \pm 0.005$}
    &\tabincell{c}{$0.855 \pm 0.003$}
    &\tabincell{c}{47.5}
    &\tabincell{c}{{66.3}}
    &\tabincell{c}{{1.96e-2}}
\\
    {\textbf{TTL-1}}
    &\tabincell{c}{\textcolor{red}{{$\mathbf{0.854 \pm 0.013}$}}}
    &\tabincell{c}{\textcolor{red}{$\mathbf{0.889 \pm 0.015}$}}
    &\tabincell{c}{\textcolor{red}{$\mb{26.11}$}}
    &\tabincell{c}{{60.17}}
    &\tabincell{c}{{1.68e-2}}
\\

\midrule

{TF-2}
    &\tabincell{c}{{{$0.849 \pm 0.020$}}}
    &\tabincell{c}{{$0.886 \pm 0.017$}}
    &\tabincell{c}{{36.1}}
    &\tabincell{c}{{64.9}}
    &\tabincell{c}{1.93e-2}
\\
{LWFT-2}
    &\tabincell{c}{{{$0.835 \pm 0.038$}}}
    &\tabincell{c}{{$0.870 \pm 0.028$}}
&\tabincell{c}{47.5}
&\tabincell{c}{66.3}
&\tabincell{c}{1.96e-2}
\\
    {\textbf{TTL-2}}
    &\tabincell{c}{\textcolor{red}{{$\mathbf{0.854 \pm 0.013}$}}}
    &\tabincell{c}{\textcolor{red}{$\mathbf{0.889 \pm 0.015}$}}
    &\tabincell{c}{\textcolor{red}{$\mb{26.11}$}}
    &\tabincell{c}{{60.17}}
    &\tabincell{c}{{1.68e-2}}
\\

\Xhline{3\arrayrulewidth}
\end{tabular}
}
\resizebox{0.45\linewidth}{!}{
\def\arraystretch{1.5}
\addtolength{\tabcolsep}{-1pt}
\begin{tabular}{c| c c c c c}
\Xhline{3\arrayrulewidth}
 Method &
Dice Coef $\uparrow$ &
Jaccard index $\uparrow$ &
P(M) $\downarrow$ &
MACs(G) $\downarrow$ &
T(ms) $\downarrow$\\
\midrule
\hline
{FTL}
& $0.968 \pm 0.029$
& $0.940 \pm \small{0.051}$
& $24.4$
& $5.93$
& $11.8$
\\
\midrule
{\textbf{TTL-B1}}
& $0.970 \pm 0.029$
& $0.941 \pm 0.052$
&\textcolor{red}{$\mb 21.3$}
&\textcolor{red}{$\mb 1.68$}
&\textcolor{red}{$\mb 7.05$}
\\
{\textbf{TTL-B2}}
&\textcolor{red}{$\mathbf{0.972 \pm 0.027}$}
&\textcolor{red}{$\mathbf{0.946 \pm 0.047}$}
&$21.5$
&$3.02$
&$8.50$
\\
{\textbf{TTL-B3}}
& $0.968 \pm 0.029$
& $0.939 \pm 0.051$
&$22.1$
&$4.82$
&$10.6$
\\
\Xhline{3\arrayrulewidth}
\end{tabular}
}

\vspace{0.3cm}
\caption{mitotic cell classification with its Grad-CAM visualization (top row), pulmonary embolism CT image classification(bottom left), lung segmentation(bottom right). The best result of each column is colored in \textcolor{red}{\textbf{red}}. $\uparrow$ indicates larger value is better and $\downarrow$ indicates lower value is better. ``-1" means with the block-wise search only, ``-2" means with the two-stage block-layer hierarchical search, ``-Bx" means model truncate at block x.}
\vspace{-0.4cm}
\label{tab:combined}
\end{table}

\noindent \textbf{COVID-19 chest x-ray image classification}\label{subsec:bimcv}
We take the chest x-rays from the BIMCV-COVID19$+$ (containing COVID positives) and BIMCV-COVID19$-$ (containing COVID negatives) datasets (iteration 1)~\citep{vaya2020bimcv}\footnote{\url{https://bimcv.cipf.es/bimcv-projects/bimcv-covid19/\#1590858128006-9e640421-6711}.}. We manually remove the small number of lateral views and outliers, leaving $2261$ positives and $2463$ negatives for our experiment. We have demonstrated the plausibility and superiority of TTL based on this MIC task around \cref{sec:ttl_method}, which we do not repeat here.  

\noindent \textbf{Mitotic cells classification}\label{subsec:mitotic}
The density of mitotic cells undergoing division (i.e., mitotic figures) is known to be related to tumor proliferation and can be used for tumor prognosis~\citep{MIDOG2022}. Since cell division changes its morphology, we expect mid-level blob-like features to be the determinant here. We take the dataset from the mitotic domain generalization challenge (MIDOG2022)~\citep{MIDOG2022} which is about detection of mitotic figures, i.e., the training set consists of properly cropped $9501$ mitotic figures and $11051$ non-mitotic figures, and the task is to localize mitotic figures on large pathological slices during the test. We modify the task into a binary MIC (positives are mitotic figures, and negatives are non-mitotic figures). 

\cref{tab:combined} (top row) summarizes the results. We observe that: \textbf{(1)} Our TTL-2 and TTL-1 beat all other TL methods, and also yield the most compact and inference-efficient models; \textbf{(2)} Both TTL-1/2 and TF find the best cutoffs at the transition of block3 and block4, implying that high-level features are possibly unnecessary and can even be hurtful for this task and confirming our tuition that mid-level visual features are likely be crucial for decision; \textbf{(3)} Of all methods, LWFT with block-wise search performs the worst; even after layer-wise search, its AUROC only matches that of the baseline FTL. The inferior performance of LWFT implies that only fine-tuning the top layers is not sufficient for this task.

\noindent \textbf{Pulmonary embolism CT image classification}\label{subsec:PE}
Pulmonary Embolism (PE) is a blockage of the blood vessels connecting the lungs and the heart, and CT pulmonary angiography (CTPA) is the gold standard for its diagnosis~\citep{WittramEtAl2004CT}. We take the public PE dataset \citep{Huang2020} consisting of $1797$ CT images from $1773$ patients, and compare the performance with handcrafted 3DCNN model, PENet which is the SOTA model that outperforms other 3D DCNN models, such as ResNet3D-50, ResNeXt3D-101, and DenseNet3D-121 as demonstrated in \citet{Huang2020}. 

On CT images, PE often appears as localized blobs that map to mid-level visual features. So the suboptimal TL performance of the SOTA models is mostly likely due to the rigid FTL strategy. We confirm this on ResNeXt3D-101 pretrained on kinetics-600: after layer-wise search, all differential TL methods including TF, LWFT, and TTL outperform PENet by considerable margins in both AUROC and AUPRC, as shown in \cref{tab:combined} (bottom left). Also, both TTL-1 and TF-1 find the best cutoff at the transition of block3 and block4, another confirmation of our intuition that probably only low- to mid-level features are needed here. We note that although the AUROC obtained via FTL is slightly lower than that of PENet, the AUPRC is actually higher---which \citep{Huang2020} does not consider when drawing their conclusion. Our TTL-2 is a clear winner in performance, despite that PENet has been meticulously designed and optimized for the task.

\noindent \textbf{Chest X-ray lung segmentation}
        \label{sec:exp_segmentation}
Despite our focus on MIC, we briefly explore the potential of TTL for segmentation also. To this end, we explore a public chest x-ray lung segmentation datasets collected from two source: Montgomery Country XCR set (MC) and Shenzhen Hospital CXR Set (SH)~\citep{jaeger2013automatic, candemir2013lung}, both of which provide manual segmentation masks. MC consists of $58$/$80$ tuberculosis/normal cases, and SH has $336$/$326$ tuberculosis/normal cases. 

For simplicity, we only perform block-wise truncation, and our experimental results are shown in \cref{tab:combined} (bottom right). TTL achieves the best segmentation performance at block 2 (TTL-B2) and outperforms FTL by $0.6\%$ in terms of Dice Coefficient and Jaccard index (both are standard metrics for evaluating segmentation performance). Notably, TTL-B2 is more efficient than FTL and reduces the model size by $12\%$ and inference time by $28\%$.

\section{Conclusion}\label{sec:conclusion}

In this paper, we present a thorough examination of transfer learning (TL) in the context of medical image classification (MIC). We introduce a novel method, named TruncatedTL (TTL), which leverages the pre-trained model's visual semantic representations and aligns them with target medical images. Through a comprehensive analysis of feature transferability, we reveal that in many MIC scenarios, low- to mid-level features suffice for effective transfer learning. We showcase four real-world medical applications, including 2D and 3D MIC (plus one 2D segmentation for exploration), to demonstrate the effectiveness of TTL. Our findings confirm that TTL significantly reduces model complexity, both in terms of size and inference speed, while achieving comparable performance to existing methods. This study contributes to the advancement of transfer learning techniques in the domain of medical imaging and underscores the potential of TTL as a promising alternative for improving model performance and efficiency in MIC tasks.

\bibliography{refs}

\newpage

\appendix

\setcounter{figure}{4}
\setcounter{table}{2}

\section{Additional experimental details}
\subsection{Experimental protocol}
\noindent\textbf{Training TL models} All 2D experiments follow the same training protocol unless otherwise stated: \textbf{1)} the given dataset is split into $64\%$ training, $16\%$ validation, and $20\%$ test; \textbf{2)} ResNet50 is the default model for all 2D MIC tasks, and ResNet34 for 2D segmentation; \textbf{3)} all 2D images are center-cropped and resized to $224\times224$; \textbf{4)} random cropping and slight random rotation are used for data augmentation during TL; \textbf{5)} the ADAM optimizer is used for all TL methods, with an initial LR $10^{-4}$ and a batch size $64$; The \texttt{ReduceLROnPlateau} scheduler in PyTorch is applied to adaptively adjust the LR: when the validation AURPC stagnates, the LR is decreased by $1/2$. Training is terminated when the LR drops below $10^{-7}$; \textbf{6)} the best model is chosen based on the validation AUPRC.

\noindent\textbf{Evaluating TL models} We measure the runtime on a system with Intel Core i9-9920X CPU and Quadro RTX 6000 GPU. We report AUROC/AUPRC for MIC, and Dice Coefficient/Jaccard Index for segmentation as performance metrics and Params, MACs, runtime time as complexity metrics.

\subsection{Extending TTL on segmentation models}
We construct the TTL model in a manner similar to what we have done for the MIC tasks. We take the U-Net architecture with ResNet34 and pretrained on ImageNet as our backbone model. To ensure the skip-connection structure can be preserved after truncation, we symmetrically truncate the backbone and segmentation head simultaneously. We do not include a comparison with LWFT and TF, as the original papers do not discuss how to extend them to segmentation and our MIC tasks above already show the superiority of our TTL. 

\begin{figure}[h]
    \centering
    \includegraphics[width=0.95\linewidth]{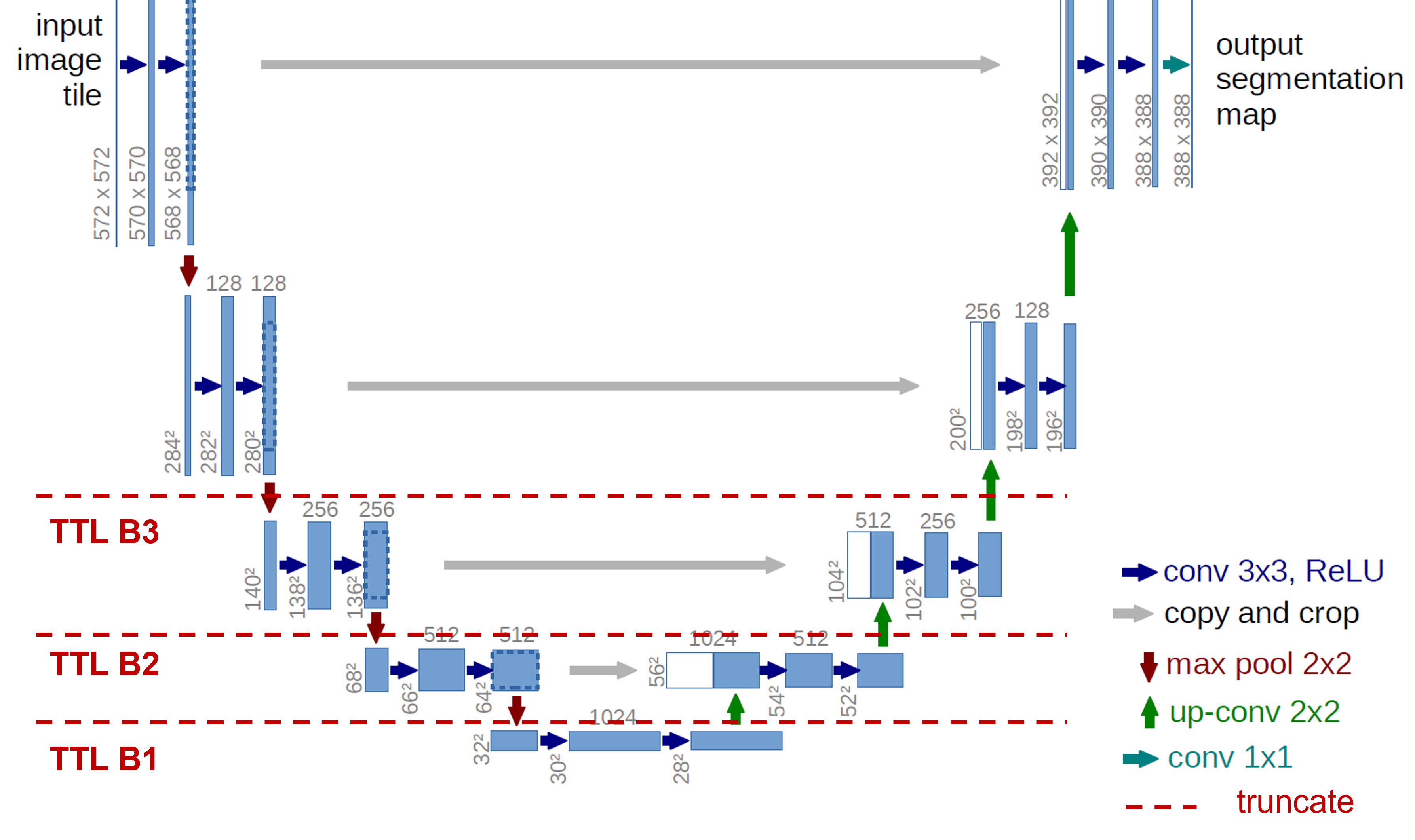}
    \caption{Similar to TTL on ResNet/DenseNet, we identify the block structure in U-Net and truncate at the intersection between blocks. The image of U-Net is adapted from~\citet{ronneberger2015u}}
    \label{fig:my_label}
\end{figure}

\section{Ablation studies}

\subsection{Impact of network architecture} 
To study the impact of network architecture on the result of TTL, we compare three network models: ResNet50, DenseNet121, and EfficientNet-b0, which represent large-, mid-,  and small-size models, respectively, on COVID-19 classification. The results are summarized in \cref{tab:arch_impact}. It is clear that our TTL uniformly improves the model performance and reduces the model size by at least several fold, compared to FTL. In particular, even for the most lightweight model EfficientNet-b0, TTL still pushes up the performance by about $5\%$ in both AUROC and AUPRC, and reduces the model size by about $4$-fold. 

\begin{table}[!htpb]
\caption{Impact of network architecture on TTL performance. The best results in each group are colored in \textcolor{red}{\textbf{red}}.}
\centering
\vspace{0.5em}
\begin{adjustbox}{width=0.9\columnwidth,center}
\label{tab:arch_impact}
\def\arraystretch{1.2}
\addtolength{\tabcolsep}{0.3em}
\resizebox{\linewidth}{!}{%
\begin{tabular}{l | l c c c c c c}
\Xhline{3\arrayrulewidth}
&
\multirow{2}[2]{*}{Method} &
\multirow{2}[2]{*}{AUROC} &
\multirow{2}[2]{*}{AUPRC} &
\multirow{2}[2]{*}{Params(M)} &
\multirow{2}[2]{*}{MACs(G)} &
\multicolumn{2}{c}{Speed(s)}\\\cline{7-8}

&
&
&
&
&
&
CPU &
GPU

\\
\midrule
\midrule
\parbox[t]{15mm}{\multirow{2}{*}{\rotatebox[origin=r]{0}{RN50$^1$}}}
&\tabincell{c}{FTL}
&\tabincell{c}{{$0.853 \pm 0.004$}}
&\tabincell{c}{{$0.861 \pm 0.002$}}
&\tabincell{c}{23.5}
&\tabincell{c}{4.12}
&\tabincell{c}{{80.0}}
&\tabincell{c}{6.72}
\\
&\tabincell{c}{TTL}
&\tabincell{c}{{\textcolor{red}{$\mathbf{0.865 \pm 0.001}$}}}
&\tabincell{c}{{\textcolor{red}{$\mathbf{0.870 \pm 0.007}$}}}
&\tabincell{c}{\textcolor{red}{\textbf{8.55}}}
&\tabincell{c}{\textcolor{red}{\textbf{3.31}}}
&\tabincell{c}{{\textcolor{red}{\textbf{60.9}}}}
&\tabincell{c}{\textcolor{red}{\textbf{5.83}}}
\\
\midrule

\parbox[t]{8mm}{\multirow{2}{*}{\rotatebox[origin=r]{0}{DN121$^2$}}}
&\tabincell{c}{FTL}
&\tabincell{c}{{$0.852 \pm 0.002$}}
&\tabincell{c}{{$0.856 \pm 0.004$}}
&\tabincell{c}{18.2}
&\tabincell{c}{3.31}
&\tabincell{c}{107}
&\tabincell{c}{24.0}
\\
&\tabincell{c}{TTL}
&\tabincell{c}{\textcolor{red}{$\mathbf{0.866 \pm 0.006}$}}
&\tabincell{c}{\textcolor{red}{$\mathbf{0.871 \pm 0.012}$}}
&\tabincell{c}{\textcolor{red}{\textbf{1.53}}}
&\tabincell{c}{\textcolor{red}{\textbf{2.11}}}
&\tabincell{c}{\textcolor{red}{\textbf{46.2}}}
&\tabincell{c}{\textcolor{red}{\textbf{5.94}}}
\\
\midrule

\parbox[t]{8mm}{\multirow{2}{*}{\rotatebox[origin=r]{0}{ENb0$^3$}}}
&\tabincell{c}{TL}
&\tabincell{c}{{$0.795 \pm 0.004$}}
&\tabincell{c}{{$0.794 \pm 0.002$}}
&\tabincell{c}{3.63}
&\tabincell{c}{0.378}
&\tabincell{c}{30.0}
&\tabincell{c}{8.33}
\\
&\tabincell{c}{TTL}
&\tabincell{c}{\textcolor{red}{$\mathbf{0.842 \pm 0.001}$}}
&\tabincell{c}{\textcolor{red}{$\mathbf{0.843 \pm 0.007}$}}
&\tabincell{c}{\textcolor{red}{\textbf{0.896}}}
&\tabincell{c}{\textcolor{red}{\textbf{0.255}}}
&\tabincell{c}{\textcolor{red}{\textbf{24.0}}}
&\tabincell{c}{\textcolor{red}{\textbf{6.46}}}
\\

\Xhline{3\arrayrulewidth}
\end{tabular}
}
\end{adjustbox}
\raggedright \footnotesize{Abbreviation: $^1$ ResNet50, $^2$:DenseNet121, $^3$ EfficientNet B0}
\end{table}

\subsection{Impact of pooling} 
\begin{figure}[h]
\begin{tabular}{cc}
\bmvaHangBox{\hspace{-2.24mm}\fbox{\includegraphics[width=6.1cm]{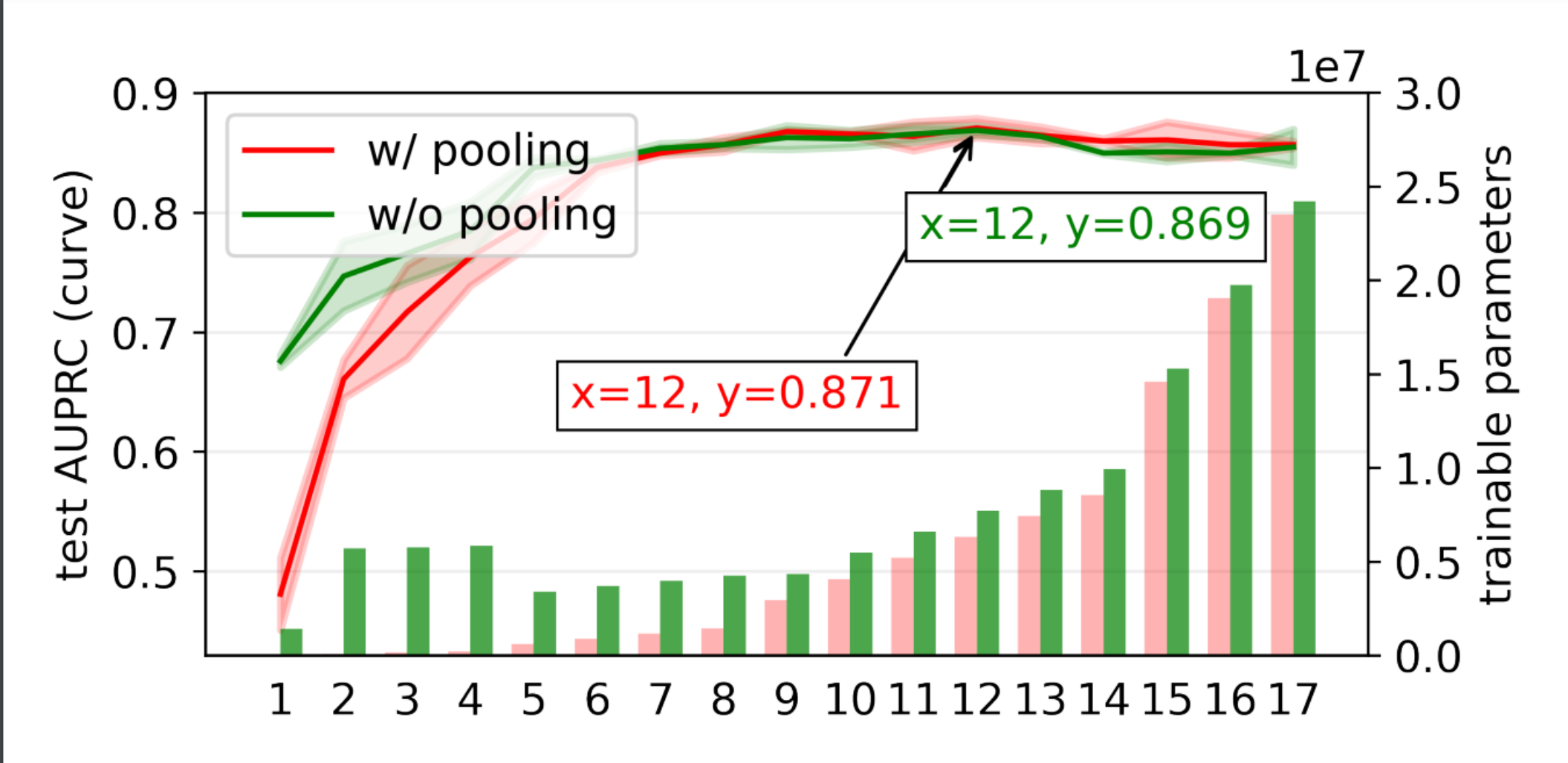}}}&
\bmvaHangBox{\hspace{-3.24mm}\fbox{\parbox{6.3cm}{\includegraphics[width=6.35cm]{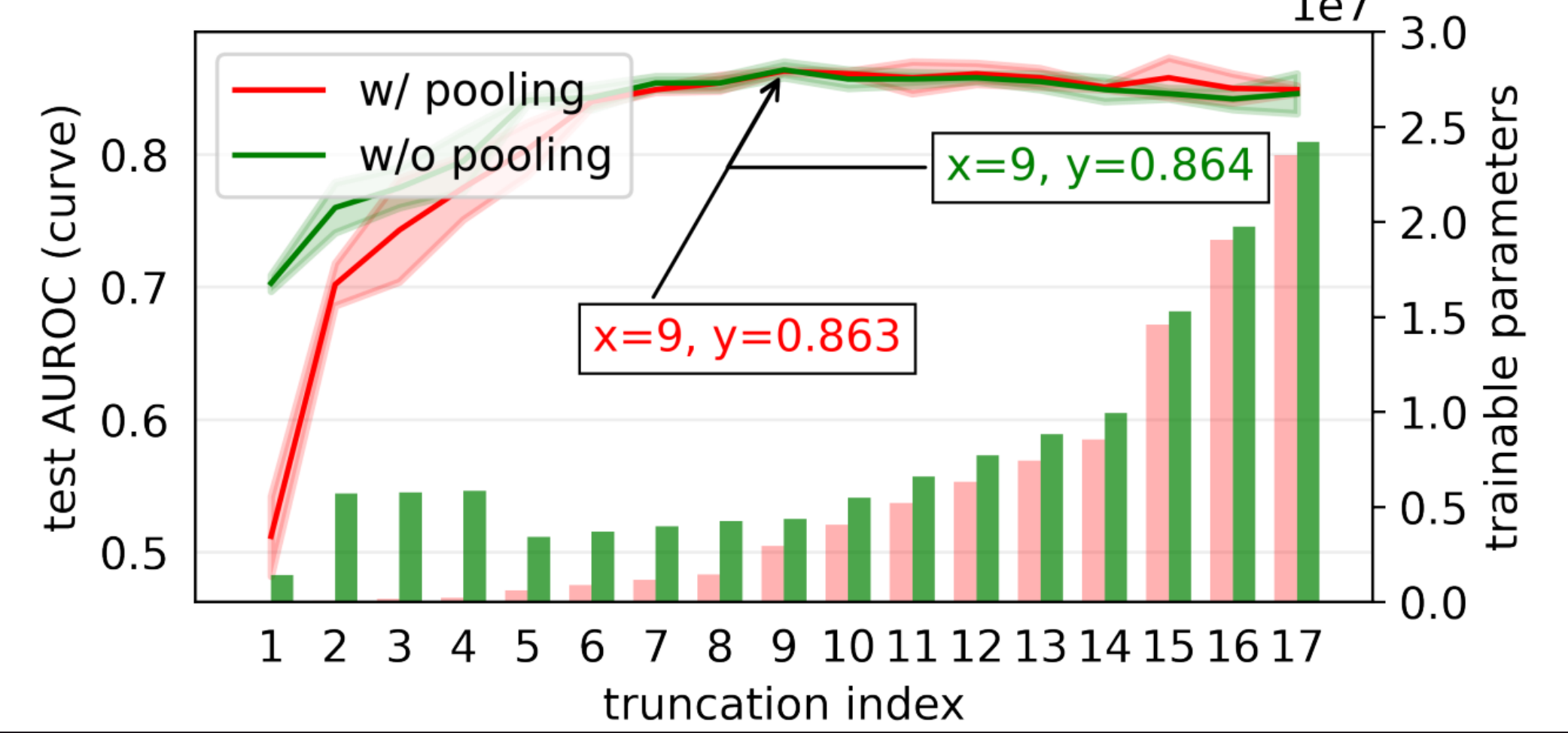}}}}
\\
(i)&(ii)
\end{tabular}
\caption{ResNet50 transferred from ImageNet to BIMCV using our TTL. We report the performance in both AURPC (i) and AUROC (ii). The red and green curves represent TTL performance with and without the extra adaptive pooling layer between the final convolution layer and the MLP classifier. Arrows point to the peak performance.}
\label{fig:pooling}
\end{figure}

After truncation, we can either directly pass the full set of features or downsample them before feeding them into the MLP classifier. For deep learning models, we can easily perform subsampling by inserting a pooling layer. For this purpose, 
we use PyTorch's built-in function \texttt{AdaptiveAvgPool2d} to downsample the feature map to $1\times1$ in the spatial dimension, which produces the most compact spatial features. We compare models with vs. without this extra pooling layer on COVID-19 classification, and present the results in \cref{fig:pooling}. We observe no significant gap between the two settings in terms of peak performance measured by both AUPRC and AUROC. However, the setting with pooling induces a lower-dimensional input to the MLP classifier, and hence contains much fewer trainable parameters compared to that without pooling.

\section{Exploratory studies}
\label{sec:exploration}

\subsection{Detecting the near-best truncation point} 
Our two-stage search strategy for good truncation points (see {\color{red}Section 3.2}) is effective as verified above, but can be costly due to the need for multiple rounds of training, each for one candidate truncation point. In this subsection, we explore an alternative strategy to cut down computation: the idea is to detect the transition point where feature reuse becomes negligible. For this, we recall the SVCCA analysis in 
{\color{red}Fig. 4}: we use the distribution of the SVCCA coefficients to quantify the correlation of features before and after FTL. To determine when the correlation becomes sufficiently small, we compare the correlation with that between random features. 

\begin{figure}[!htbp]
\centering 
\includegraphics[width=0.5\linewidth]{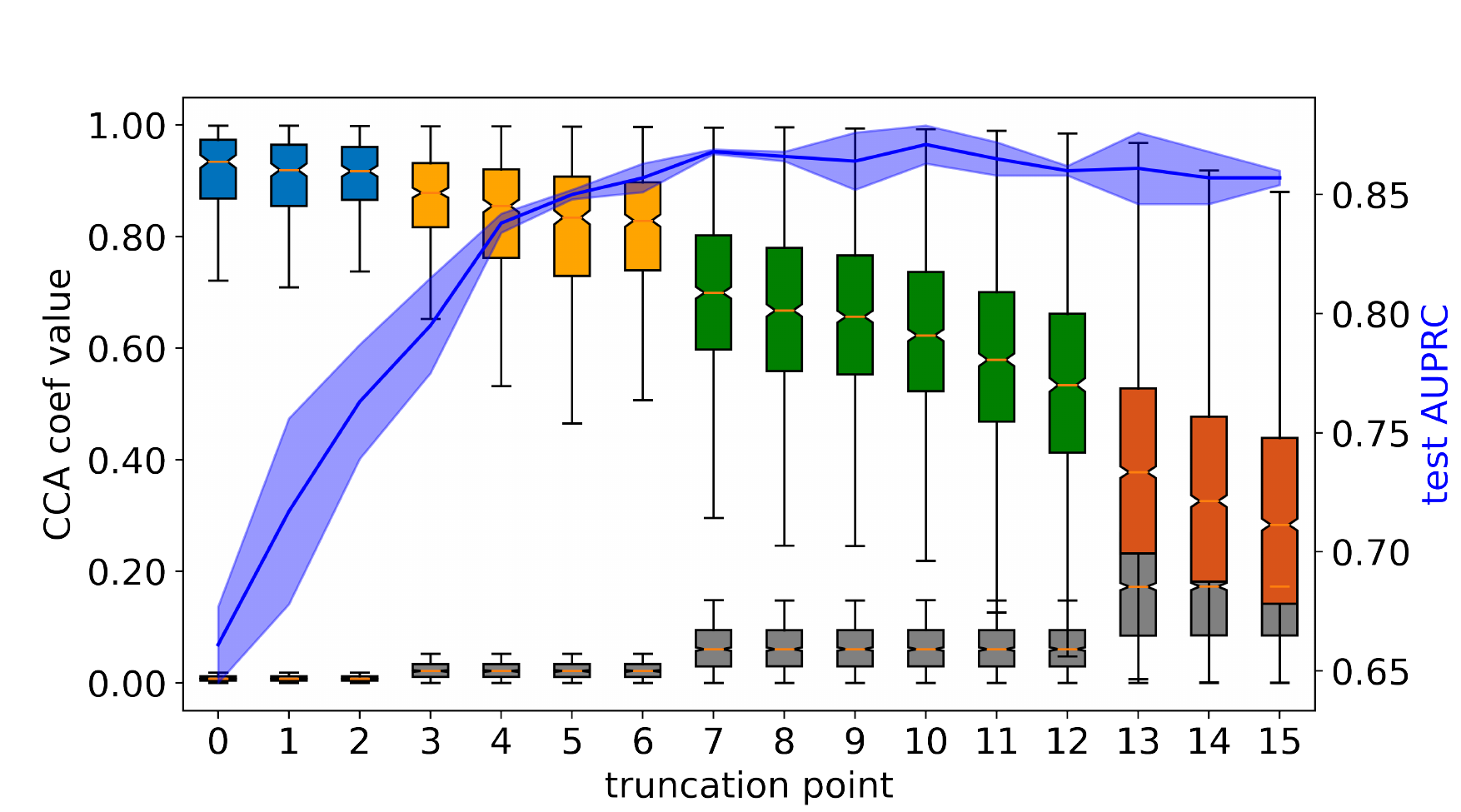}
\includegraphics[width=0.48\linewidth]{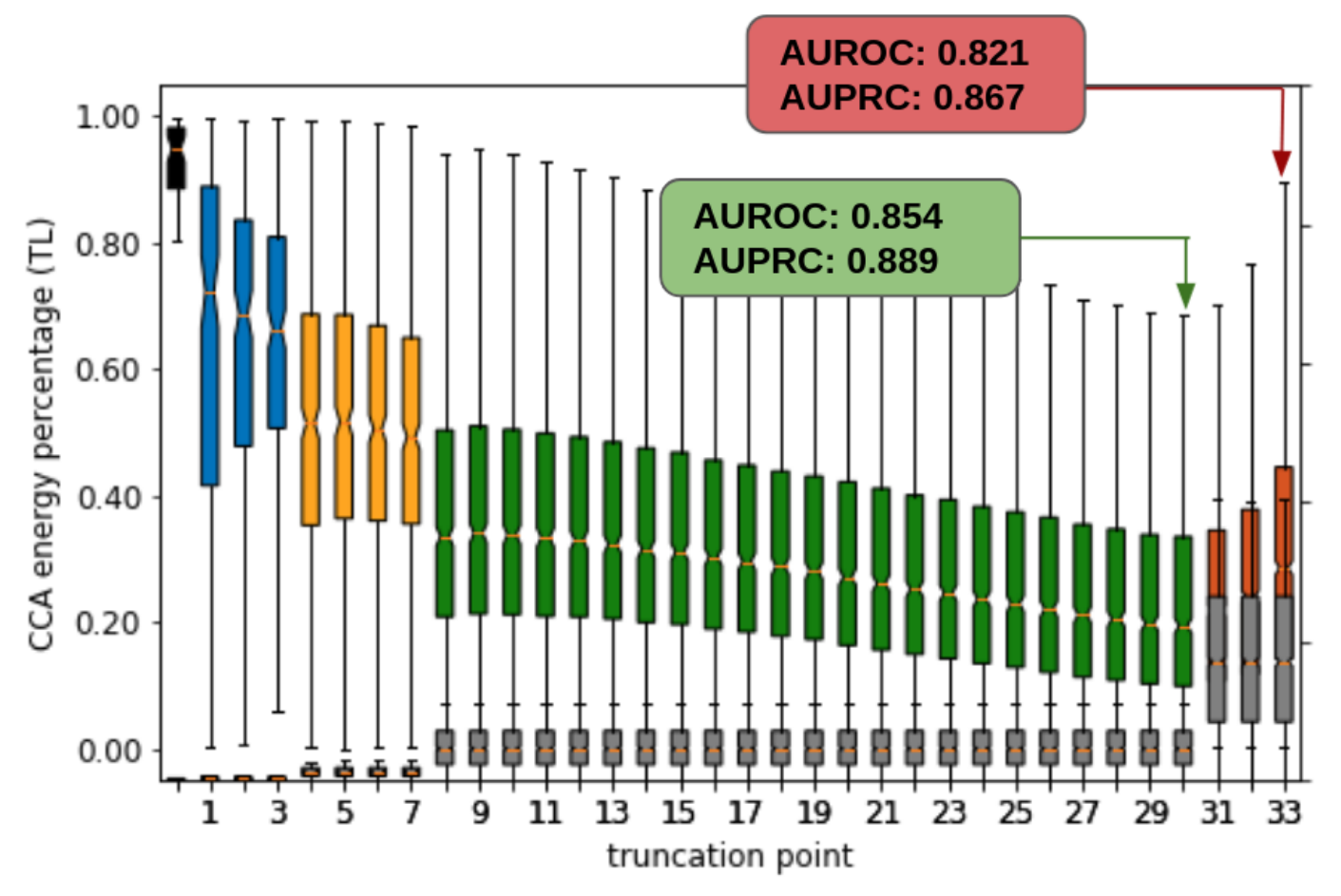}
\caption{\textbf{Left}: SVCCA on ResNet50 transferred from ImageNet to BIMCV. The plot is inherited from {\color{red}Fig. 3} (ii). The gray boxes and the associated bars represent the distributions of CCA coefficients between random features. \textbf{Right}: SVCCA on ResNeXt3D-101 transferred from Knetic-600 to CTPA. Truncating at layer 30 (marked in green) significantly outperforms FTL (marked in red).}
\label{fig:svcca_bimcv}
\end{figure}

We quickly test the new strategy on both COVID-19 and CT-based PE classification, shown in \cref{fig:svcca_bimcv} left and \cref{fig:svcca_bimcv} right, respectively. From \cref{fig:svcca_bimcv}, we can see that for COVID-19 classification, feature reuse diminished after the $12$-th candidate truncation point: from $13$-th onward, feature correlation becomes very close to that between random features, suggesting that the features are almost uncorrelated. As expected, the $12$-th truncation point also achieves near-optimal performance as measured by AUPRC. Similarly, for CT-based PE classification in \cref{fig:svcca_bimcv} right, this strategy suggests the $30$-th truncation point, which yields the same result as that we have obtained using the two-stage search strategy as reported in {\color{red}Table 2} bottom left, in both AUPRC and AUROC. Note that in \cref{fig:svcca_bimcv} right, the SVCCA coefficients slightly increase after the $30$-th truncation point due to dimensionality: the channel sizes are doubled there compared to the previous blocks, and the raised dimension induces higher correlation coefficients. But to decide if feature reuse becomes trivial, we only need to check if the two SVCCA-coefficient distributions are sufficiently close, not their absolute scalings. 

Thus, this new strategy seems promising as a low-cost alternative to our default two-stage search strategy. We summarize the key steps as follows. 

\textbf{Step 1}: perform FTL on the target dataset; 

\textbf{Step 2}: compute SVCCA coefficients between the features before and after FTL, at all candidate truncation points; 

\textbf{Step 3}: compute SVCCA coefficients between random features,  at all candidate truncation points; 

\textbf{Step 4}: locate the truncation point where the distribution of the two groups of SVCCA coefficients from \textbf{Step 2} and \textbf{Step 3} becomes substantially overlapped, and take it as the truncation point; 

\textbf{Step 5}: fine-tune the model with the truncation point selected from \textbf{Step 4}. 

Compared to our two-stage strategy that entails numerous rounds of model finetuning, the new strategy only requires two rounds of finetuning: one from \textbf{Step 1} on the whole model, and the other from \textbf{Step 5} on the truncated model. 
\begin{table}[!htbp]
\centering
\caption{Symptom of lung disease in CheXpert}
\begin{adjustbox}{width=0.7\columnwidth,center}
\def\arraystretch{1.2}
\addtolength{\tabcolsep}{0.5em}
\label{tab:related_work}
\begin{tabular}{l|lrl|l}
\Xhline{3\arrayrulewidth}
Diseases & Shape & Edge & Contrast  & level of features  \\ \hline
No Finding  &\tobedone &\tobedone &\tobedone & None \\
Enlarged Cardiom.  &\done &\tobedone &\done & low and high\\
Cardiomegaly   &\done   &\tobedone &\done & low and high\\
Lung Opacity    &\tobedone &\tobedone &\done    & low\\
Lung Lesion     &\done  &\tobedone  &\done  & low and high\\
Edema   &\tobedone  &\tobedone  &\done  & low and high\\
Consolidation &\tobedone    &\tobedone  &\done & low\\
Pneumonia   &\tobedone  &\tobedone  &\done  & low\\
Atelectasis &\done  &\tobedone  &\done  & low and high\\
Pneumonthorax   &\tobedone  &\done  &\done  &low\\
Pleural Effusion    &\tobedone  &\tobedone  &\done  &low\\
Pleural Other   &\tobedone  &\tobedone  &\done  &low\\
Fracture    &\done  &\done  &\done  &low and high\\
Support device  &\done  &\done  &\done  & low and high\\
\Xhline{3\arrayrulewidth}
\end{tabular}
\label{tab:chexpert_disease_symptom}
\end{adjustbox}
\end{table}
\begin{figure}[!htbp]
\centering
\hspace{-2.24mm}\includegraphics[width=0.7\linewidth]{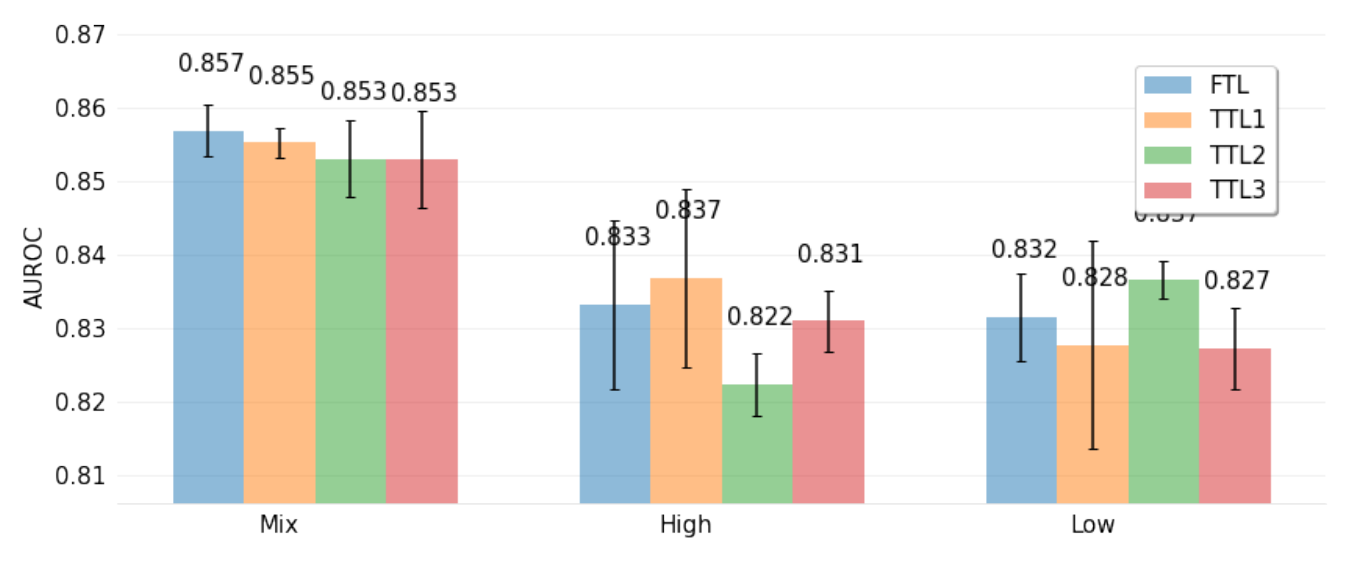}\\
\includegraphics[width=0.70\linewidth]{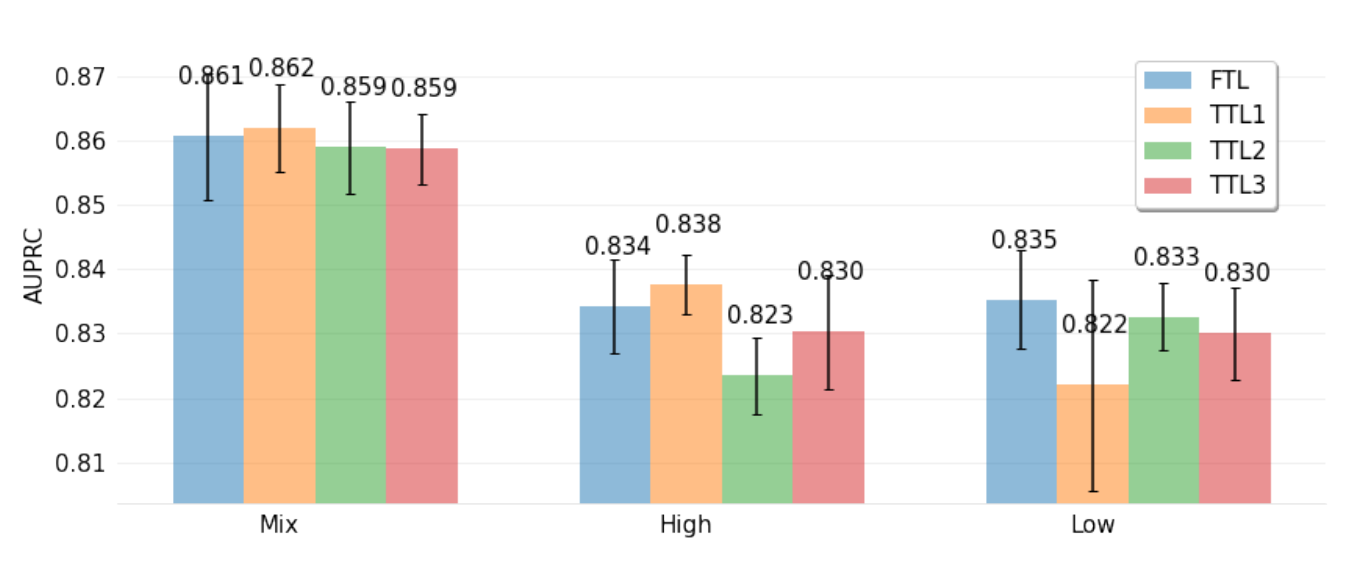}\hspace{-2.24mm}
\caption{ResNet50 transferred from CheXpert to BIMCV. (top) test AUROC score; (bottom) test AUPRC score. ``Mix", ``High", and ``Low" mean finetuning from pretrained models on the original CheXpert, CheXpert-high, and CheXpert-low, respectively. TTLx means the truncation point is chosen at the transition point between block x and block x+1.}
\label{fig:med2med}
\end{figure}%
\subsection{Inter-domain TL}
All we have discussed above are intra-domain TL where the source domain is distinct from the target domain. It is tempting to think inter-domain TL might be more effective. To test this, we take the x-ray-based COVID-19 classification task again, but now finetune the ResNet50 model pretrained on CheXpert, a massive-scale x-ray dataset ($\sim 220K$ images) for disease prediction covering $13$ types of diseases (see \cref{tab:chexpert_disease_symptom})~\citep{irvin2019chexpert}. These diseases correspond to varying levels of visual features. Hence, we categorize them into two groups: \textbf{CheXpert-low} includes diseases that need low-level features only, and \textbf{CheXpert-high} covers those needing both low- and high-level features. A summary of the categorization can be found in \cref{tab:chexpert_disease_symptom}\footnote{{Disease symptoms information obtained from Mayo clinic: \url{https://www.mayoclinic.org/symptom-checker/select-symptom/itt-20009075}}}. We pretrain ResNet50 on $3$ variants of the dataset, respectively: full CheXpert, CheXpert-low, and CheXpert-high, and then compare FTL and our TTL on the three resulting models. For TTL, we only perform a coarse-scale search and pick the $3$ transition layers between the $4$ blocks in ResNet50 as truncation points. As always, all experiments are repeated three times. 

From the results summarized in \cref{fig:med2med}, we find that: \textbf{(1)} Our TTL always achieves comparable or superior performance compared to FTL, reaffirming our conclusion above; \textbf{(2)} Transferring from the model pretrained on the original dataset substantially outperforms transferring from those pretrained on the CheXpert-high and CheXpert-low subsets. This can be explained by the diversity of feature levels learned during pretraining: on the subsets more specialized features are learned; in particular, on CheXpert-low, perhaps only relatively low-level features are learned; and \textbf{(3)} Notably, our results show that medical-to-medical TL does not do better than TL from natural images, when we compare the results in \cref{fig:med2med} with those in {\color{red}Table 1}. We suspect this is because feature diversity is the most crucial quality required on the pretrained models in TL, and models pretrained on natural images perhaps already learn sufficiently diverse visual features.

\section{Sample images of Grad-CAM analysis}
To find out what visual features are learned by different models, we conduct a feature analysis study using Grad-CAM\citep{selvaraju2017grad}. Our empirical findings show that deep models tend to produce spatially extensive features that spread over the image. In contrast, shallow models such as those produced by TTL tend to find spatially localized features that often capture the infected area around the lungs and the mitotic region, as shown in Fig. 1 and Table 2. In particular, those localized regions often take the form of the abnormal texture of boundary changes and are invariant to the global object (e.g., the shape of the ribs and cells), which demonstrates the effectiveness of TTL in learning more discriminative features for MIC.

\section{Feature visualization on pretrained CNN}\label{sec:feature_vis}
\begin{figure}[h]
    \centering
    \hspace{-5.24mm}\includegraphics[width=0.95\linewidth]{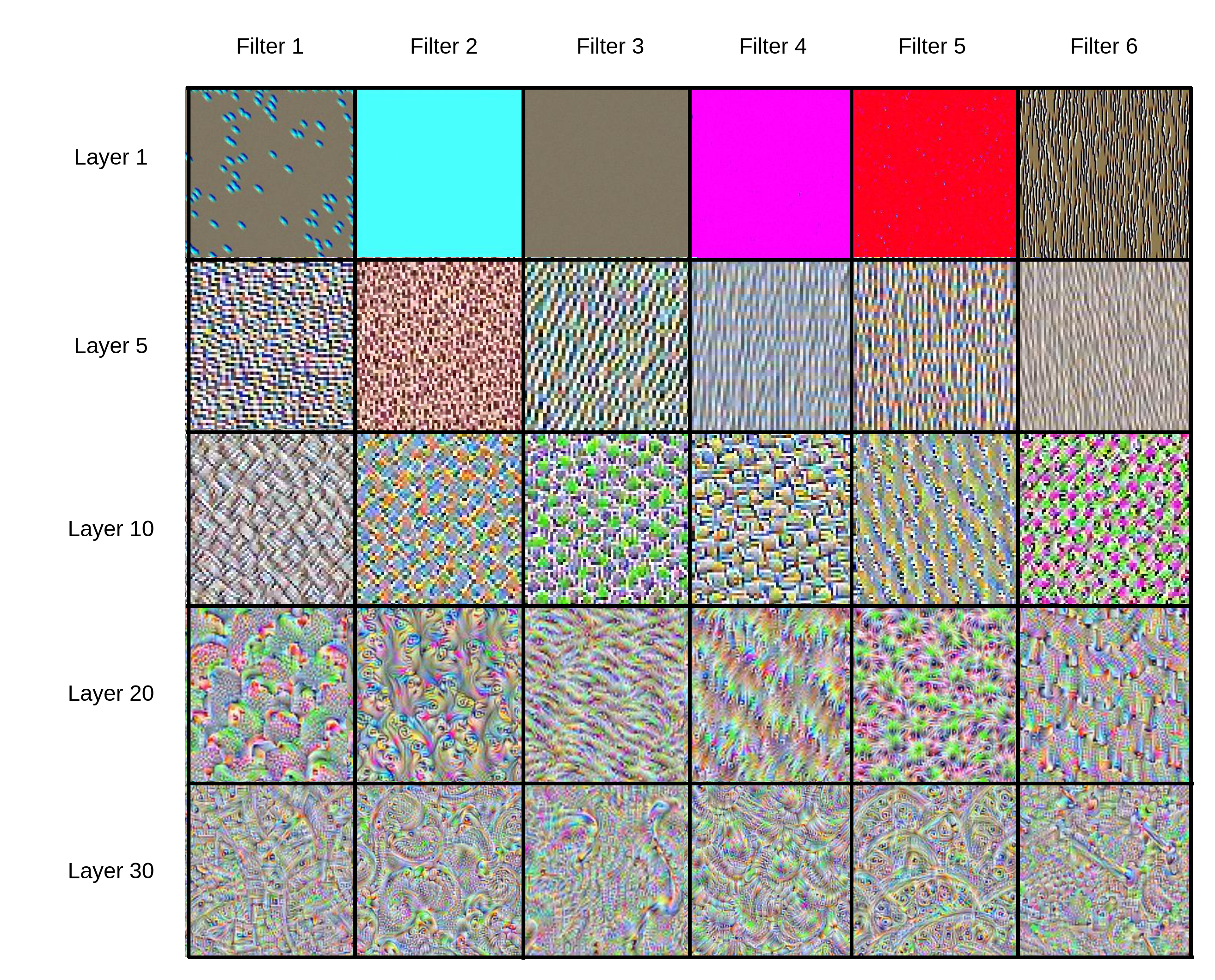}
    \caption{Visualized features learnt at different layers in VGG19 pretrained on ImageNet. To get the visualized images, we optimize the inputs with respect to the output value from a certain filter.}
    \label{fig:filter_vis}
\end{figure}

To gain insights into the features acquired by layers in a CNN pre-trained on a general natural image dataset, we employed visualization techniques to elucidate the learned features across various layers in a VGG19 network pretrained on ImageNet. The results are shown in \cref{fig:filter_vis}. Our empirical observation reveals that bottom layers (e.g., layer 1, 5, and 10) mostly learn low-semantic patterns such as color and simple structured patterns. As delve deeper into top layers (e.g., layer 20 and 30), it progressively refines its knowledge, capturing higher-level semantic characteristics, including specific object-related local regions.

\end{document}